\documentclass{svjour3}                     
\smartqed  
\usepackage{amsmath}
\usepackage{amssymb}
\usepackage{enumerate}
\usepackage{eucal}
\usepackage{eufrak}
\usepackage{bm}
\usepackage{graphicx}
\usepackage{algorithm}
\usepackage{algpseudocode}
\usepackage[usenames,dvipsnames]{color}
\usepackage{hyperref}
\usepackage{mathptmx}      
\newcommand{\N}{\mathbb{N}}

\newcommand{\R}{\mathbb{R}}

\newcommand{\Prob}{\mathbf{P}}
\newcommand{\Expec}{\mathbf{E}}

\newcommand{\bigO}{\bm{O}}

\newcommand{\ee}{\mathrm{e}}
\newcommand{\transp}{\mathrm{t}}

\newcommand{\etal}{\emph{et al.}}
   
\newcommand{\ignore}[1]{}

\makeatletter
\def\BState{\State\hskip-\ALG@thistlm}
\makeatother

\begin{document}

\title{Stochastic Modeling and Simulation of Viral Evolution}



\author{
Luiza Guimar\~aes
\and
Diogo Castro 
\and \\
Bruno Gorzoni 
\and
Luiz Mario Ramos Janini
\and
Fernando Antoneli
}

\authorrunning{Guimar\~aes \etal} 

\institute{
Luiza Guimar\~aes 
\and
Diogo Castro
\and
Bruno Gorzoni \at
Programa de P\'os-Gradua\c{c}\~ao em Infectologia,
Universidade Federal de S\~ao Paulo, S\~ao Paulo, SP, Brazil.
\and
Luiz Mario Ramos Janini \at
Departamentos de Microbiologia, Imunologia, Parasitologia and Medicina,
Laborat\'orio de Retrovirologia, \\
Universidade Federal de S\~ao Paulo, S\~ao Paulo, SP, Brazil.
\and
Fernando Antoneli \at
Departamento de Inform\'atica em Sa\'ude,
Laborat\'orio de Biocomplexidade e Gen\^omica Evolutiva, \\
Universidade Federal de S\~ao Paulo, S\~ao Paulo, SP, Brazil. \\
\textbf{Corresponding author.} \email{fernando.antoneli@unifesp.br}
}

\date{\today}

\maketitle

\begin{abstract}
RNA viruses comprise vast populations of closely related, but highly genetically diverse, 
entities known as \emph{quasispecies}. 
Understanding the mechanisms by which this extreme diversity is generated and maintained 
is fundamental when approaching viral persistence and pathobiology in infected hosts.  
In this paper we access quasispecies theory through a mathematical model based on the
theory of multi-type branching processes, to better understand the roles of mechanisms
resulting in viral diversity, persistence and extinction. 
We accomplish this understanding by a combination of computational simulations and the
theoretical analysis of the model. 
In order to perform the simulations we have implemented the mathematical model into a 
computational platform capable of running simulations and presenting the results in a 
graphical format in real time. 
Among other things, we show that the establishment of virus populations may display four
distinct regimes from its introduction into new hosts until achieving equilibrium or 
undergoing extinction. 
Also, we were able to simulate different \emph{fitness distributions} representing 
distinct environments within a host which could either be favorable or hostile to the 
viral success. 
We addressed the most used mechanisms for explaining the extinction of RNA virus 
populations called \emph{lethal mutagenesis} and \emph{mutational meltdown}.
We were able to demonstrate a correspondence between these two mechanisms implying the 
existence of a unifying principle leading to the extinction of RNA viruses.

\end{abstract}

\section{Introduction}
\label{sec:INTRO}

Viruses with RNA genomes, the most abundant group of human pathogens~\cite{DH97}, 
exhibit high mutational rates, fast replicative kinetics, large population sizes, 
and high genetic diversity.
Current evidences also indicate that RNA virus populations consist of a wide and
interrelated distribution of variants, which can display complex evolutionary dynamics.
The complex evolutionary properties of RNA virus populations features the modulation 
of viral phenotypic traits, the interplay between host and viral factors, and other 
emergent properties~\cite{D85,DMP06}.
During viral infections, these features allow viral populations to escape from host
pressures represented by the actions from the immune system, from vaccines 
and to develop resistance antiviral drugs.
Taken together these features represent the major obstacle for the success and 
implementation of effective therapeutic intervention strategies.

In order o describe the evolution of RNA viruses and its relationship with their hosts 
and antiviral therapies, theoretical models of virus evolution have been developed.
These models employ mathematical and computational tools as methodological instruments
allowing one to address evolutionary questions from a different perspective than the 
commonly seen use of modern experimental technologies.
This kind of approach allows the implementation of low-cost research projects addressing 
evolutionary questions that are usually investigated by experimental methods.
At a deeper level, they provide a systematic perspective of the biological phenomenon, 
when viewed as \emph{proof-of-concept models}~\cite{SBD14}.
Verbal or pictorial models have long been used in evolutionary biology to formulate
abstract hypotheses about processes and mechanisms that operate among diverse species
and across vast time scales.
Used in many fields, proof-of-concept-models test the validity of verbal or pictorial 
models by laying out the underlying assumptions in a mathematical framework.

Eigen and Schuster~\cite{E71,ES79} proposed and analyzed a deterministic model for 
the evolution of polynucleotides in a dialysis reactor based on a system of ordinary, 
differential equations called \emph{quasispecies model}.
Subsequently, Demetrius \etal~\cite{DSS85} proposed a stochastic quasispecies model
in order to overcome some drawbacks of the deterministic quasispecies model of Eigen 
and Schuster~\cite{ES79}.
The approach of Demetrius \etal~\cite{DSS85} employed very powerful methods based 
on the theory of stochastic branching processes. 
This theory, originally developed to deal with the extinction of family names 
(Watson and Galton~\cite{WG1874}), has been applied since the forties to a great 
variety of physical and biological problems~\cite{H63,AN72,K02}.
On the experimental side, an early study of the RNA phage $\mathrm{Q\beta}$ reporting 
that sequence variation in a population was high but approximately stable over time 
around a consensus sequence, gave the initial stimulus to consider the notion of 
quasispecies in the broader context of RNA viruses~\cite{DSTW78}.

Since then, quasispecies theory has been recognized as a subset of theoretical
population genetics~\cite{W05,TH07}.
Recently, in the series of papers~\cite{C15a,C15b,D15,CD16}, it has been rigorously 
shown that the Wright-Fisher and Moran models for multi-loci mutation-selection  
converges to the single-peak fitness landscape quasispecies model, in the appropriate 
limit of infinite populations.
Moreover, due to is a capability to accommodate high mutation rates, it has been 
widely applied to model the evolution of viruses with RNA genomes~\cite{E93}.

Inspired by the stochastic quasispecies model of Demetrius \etal~\cite{DSS85} and 
based on branching process techniques, Antoneli \etal~\cite{ABCJ13b,ABCJ13a} 
proposed a mathematical model aimed at understanding the basic mechanisms and
phenomena of the evolution of highly-mutating viral populations replicating in a 
single host organism, called \emph{phenotypic (quasispecies) model}.
It is denominated ``phenotypic'' due to the fact that it only comprises probabilities
associated with the occurrence of deleterious, beneficial and neutral effects that 
operate directly on the replicative capability of viral particles, without any 
explicit reference to their genome.
In~\cite{D16} Dalmau introduced another generalization of the stochastic quasispecies
model also based on multitype branching processes but retaining the genotypic 
character of Demetrius \etal~\cite{DSS85}.

The phenotypic model~\cite{ABCJ13b,ABCJ13a} is defined through a probability generating 
function which formally determines the transition structure of the process.
The matrix of first moments of the branching process, or simply the \emph{mean matrix}, 
defines a deterministic linear system which describes the time evolution of conditional
expectations, a ``mean field model'' for the actual stochastic process which is 
equivalent to the \emph{Eigen's selection equation}~\cite{DSS85}. 
The deterministic mean field model has been studied by several researches, 
but without the connection to a stochastic branching process, 
see for instance~\cite{BMR99,MLPED03,C11}.

As shown in~\cite{ABCJ13b,ABCJ13a}, the phenotypic model is fully specified by three 
fundamental parameters: the probabilities of occurrence of deleterious and beneficial 
effects $d$ and $b$ -- the probability of occurrence of neutral effects is fixed by 
the complementary relation $c=1-d-b$ -- and the maximum replicative capability $R$.
By an exhaustive analysis of this ``parameter space'' we were able to depict a fairly
detailed portrait of all possible behaviors of the model.
In~\cite{ABCJ13b} we carry out a thorough analysis of mean matrix, assuming that
beneficial effects are absent and were able to show that the phenotypic model is 
``exactly solvable'', in the sense that the spectral problem for the mean matrix 
has an explicit solution.
In~\cite{ABCJ13a} we employ \emph{spectral perturbation theory} in order to treat 
the general case of small beneficial effects.
This approach has provided a complete description of the generic behavior of the model.

In the present paper, we further address the biological implications of modeling RNA 
virus populations in terms of the phenotypic model.
We achieved this goal by a combination of computational simulations and basic results 
of the theory of multitype branching processes as used in~\cite{ABCJ13b,ABCJ13a}.
In order to perform the simulations we have implemented the phenotypic model into a 
computational platform capable of running the simulation and presenting the results 
in graphical format in real time.

We start with the description of the computational platform (its interface, output and 
main simulation routine). 
Then we proceed to use some of the theoretical results of~\cite{ABCJ13b,ABCJ13a} to 
validate the program with several simulation experiments that can be read independently 
from each other and are used to evaluate distinct features of the program.
Finally, we perform two additional simulation experiments to address the main 
questions of this paper:
\begin{enumerate}[(1)]
\item What is the impact of fitness distributions on the evolution of the phenotypic 
      model and how to measure it?
\item Is there an extinction mechanism similar to the ``mutational meltdown''
      in the phenotypic model?
\end{enumerate}

\paragraph{Role of fitness distributions.} The fitness distributions of the phenotypic 
model are discrete distributions forming location-scale families parameterized by 
the replicative classes that control the progeny sizes at each replication cycle.
They can be seen as representing distinct ``compartments'' in the host which can be 
more favorable or pose restrictions to the viral replication process. 
For instance, some distributions have a positive influence on the replication, 
by enhancing the replication of particles in the higher replicative classes,
while other distributions have an opposite effect.
Examples of favorable compartments would be sites associated with immune privilege, 
or with lower concentration of antiviral drugs, or allowing for cell to cell virus 
transmission. 
Unfavorable compartments are sites with high antiviral drug penetration, small 
number of target cells, or accessed by elements of host responses as antibodies, 
citotoxic cells and others.
In this sense, we may think of fitness distributions as an environmental component 
during viral evolution.
We show that the impact of the fitness distributions on the branching process is 
subtle and can not be detected by quantities that depend only on the first moments 
of the process.
Nevertheless, we introduce a new quantity, called \emph{populational variance}, 
that is capable to detect the influence of different fitness distributions and 
is analytically and computationally tractable.

\paragraph{Unifying principle for extinction.} According to the phenotypic model a
virus population can be become extinct or eradicated from the host by the fulfillment
of a condition involving only the probability of occurrence of deleterious effects
$d$ and the maximum replicative capability $R$.
Even further, in the absence of beneficial or compensatory effects, the fate of the 
population determined by the product $R(1-d)$. If it is greater than $1$ the population
will survive or if it is lesser than $1$ the population will face extinction.
Based on this result we show that there is a correspondence between two well known
distinct mechanisms of extinction:
\begin{enumerate}[(1)]
\item \emph{Lethal Mutagenesis~\cite{LEKZRM99,BSW07,BSW08}.} The process of extinction
      of the viral population due to the increment of the deleterious rate.      
\item \emph{Mutational Meltdown~\cite{LBBG93,LG90}.} The process of extinction of the
      viral population through the step-wise loss of the fittest replicative classes
      due to \emph{random drift} associated with the finite population size effect.
\end{enumerate}
The correspondence between the two mechanisms reinforces the view that both
are ``two sides of the same coin''~\cite{MOBJ17}.
We propose here an
\emph{unifying principle for the extinction of a virus population}:
This principle is based on a mathematical model containing probabilities of
neutral and deleterious effects and the average growth rate or average maximum
fitenss which is equivalent (under appropriate interpretation) to the extinction
threshold of a branching process given by the malthusian parameter. 
In the course of the proof (see Section~\ref{subsec:FPSMM}) we consider another
parameter present in our phenotypic model, called the \emph{carrying capacity}.
Initially, it was introduced as a convenient step for the computational
implementation of the model, i.e., to prevent the population to grow boundlessly.
Nevertheless, it can be seen as a genuine parameter of the model, which controls
the intensity of the random drift. Because of this, we may consider our model as a
\emph{self-regulated branching process}, instead of a ``pure'' branching process.
Furthermore, we observe that, even though the extinction mechanisms have the same 
mathematical ``origin'', the processes leading to the actual extinction of the viral 
population may display distinct ``signatures''.

\paragraph{Structure of the paper.} The paper is structured as follows.
In section \ref{sec:SPM} we introduce the computational platform for the simulation 
of the phenotypic model.
In section \ref{sec:CPM} we perform several simulation experiments to validate the
program by comparing its output with the theoretical results from~\cite{ABCJ13b,ABCJ13a}.
We end this section with the presentation of the new results on the role of the fitness
distributions and the mutational meltdown.
The validation subsections and the two subsections on new results depend only on section
\ref{sec:SPM} and so can be read independently from each other.
The paper ends with a conclusion section.
There are 5 appendices.
Appendices A, B and C provide some background on branching process theory and theoretical 
results about the phenotypic model for the reader's convenience.
Appendices D and E provide some details about the implementation of the computational
platform introduced in the paper.

\section{Software Description}
\label{sec:SPM}

In this section introduce a computational platform for the simulation of the phenotypic 
model of~\cite{ABCJ13b,ABCJ13a}.

\subsection{The ENVELOPE Program}

The \texttt{ENVELOPE} (EvolutioN of Virus populations modELd by stOchastic ProcEss)
program is a cross-platform application developed to simulate the phenotypic model
of~\cite{ABCJ13b,ABCJ13a}.
The software contains a graphical interface to input data, visualize graphics in 
real time, and export the output data to \texttt{CSV} format, which can be used 
with a wide range of statistical analysis tools. 
It was written in \texttt{C++} programming language using the \texttt{Qt} framework 
to design the graphical user interface. 
It was exhaustively tested on \texttt{Linux} operating systems.

\begin{figure}[ht] 
\begin{center}
 \includegraphics[scale=1.6]{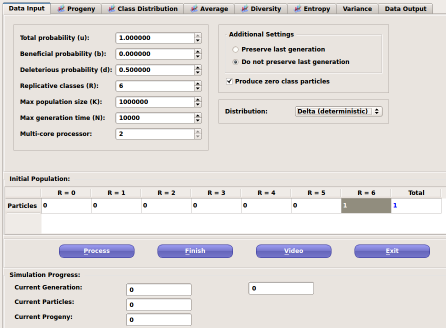}
 \caption{\label{fig:input} The ``Data Input'' tab of the \texttt{ENVELOPE} program.} 
\end{center}
\end{figure}

The main window of the program has several tabs with the first called ``Data Input''
where the user can set the values of several parameters that completely specify the 
model, as follows (see Figure~\ref{fig:input}).

\begin{itemize}
\item \emph{Total probability} ($u$): the probability that a progeny particle
       will undergo some fitness effect. It should be a number between $0$ and $1$.
       The effect of this probability is to renormalize the other probabilities 
       ($p\mapsto u\,p$) and its default value is $u=1$ (no renormalization).
\item \emph{Beneficial probability} ($b$): the probability of occurrence of a beneficial 
       effect. It should be a number between $0$ and $1$. 
\item \emph{Deleterious probability} ($d$): the probability of occurrence of deleterious 
       effect. It should be a number between $0$ and $1$. \\[1mm]
       The complementary probability $c=1-b-d$ is the probability of occurrence of
       neutral effect. If $b+d>1$ then $c$ is set to $0$ and $d=1-b$. 
\item \emph{Replicative classes} ($R$): the number of non-zero replicative classes, 
       hence there are $R+1$ replicative classes (maximum replicative capability).
\item \emph{Max population size} ($K$): the maximum population size (carrying capacity).
\item \emph{Maximum generation time} ($N$): the total number of generations to be 
      simulated.
      Each generation corresponds to a replication cycle.
\item \emph{Multi-core processor}: controls the recruitment of processors by the program.
\item \emph{Initial population}: the number of particles in each replicative class that 
      will initiate the process.
\item \emph{Distribution}: location-scale family of fitness distributions 
      (see Table~\ref{tab:TAB3}).
\end{itemize}

\renewcommand{\arraystretch}{2}

\begin{table}[ht]
\begin{center}
\begin{tabular}{c@{\qquad}c@{\qquad}c} \hline     
 Distribution & Family ($r\geqslant 1$) & Variance \\ \hline
 Deterministic & $t_r(k)=\delta_{rk}$ & $0$ \\
 Poisson & $t_r(k)=\ee^{-r}\tfrac{r^k}{k!}$ & $r$ \\
 Geometric & $t_r(k)=\tfrac{1}{r+1}\big(1-\tfrac{1}{r+1}\big)^{k}$ & $r(r+1)$ \\
 $\tfrac{1}{2}$-Binomial & $t_r(k)=\binom{2r}{k} \tfrac{1}{2^r}$ & $r/2$ \\
 Power law & $t_r(k)=\mathfrak{z}_r(k)$ & $+\infty$ \\
\hline
\end{tabular}
\end{center}
\caption{\label{tab:TAB3} Location-scale families of fitness distributions.
 All distributions are normalized so that the expectation value of $t_r$ is $r$
 and $t_0(k)=\delta_{0k}$.
 See Appendix~\ref{sec:APP0} for the notation $\delta_{rk}$ and 
 Appendix~\ref{sec:APP2} for the definition of the family of power law 
 distributions $\mathfrak{z}_r(k)$.}
\end{table}

\renewcommand{\arraystretch}{1}

The remaining tabs (``Progeny'', ``Class Distribution'', ``Average'', ``Diversity'',
``Entropy'', ``Variance'') display graphics of the above quantities in real time as 
the simulation proceeds.
The tab ``Data Output'' displays a table with all the data generated during the
simulation.
This data can be saved to a file (button ``Save to File'') or copied to the memory 
(button ``Copy to Memory'') and then it can be directly pasted into a spreadsheet.

The button ``Process'' starts the simulation, the button ``Finish'' ends 
the simulation at any time and the button ``Exit'' closes the program.
If the total number of particles in a generation is equal to zero, it is 
assumed that the population has become extinct and hence the simulation stops.
The button ``Video'' pauses the simulation, without ending the simulation, 
and allows the user to change the above parameter settings and continue 
the simulation with the new setting.
This feature is used to emulate the changes in the environment -- the host 
organism -- where the reproduction process takes place.

The evolution of the population can be measured through a few simple 
quantities that vary as a function of the generation number $n \geqslant 0$.
Let $\bm{Z}_n=(Z_n^0,\ldots,Z_n^R)$ denote the vector whose component $Z^r_n$ 
is the number of particles in the $r$-th replicative class at generation $n$.
\begin{itemize}
\item \emph{Progeny size}: total number of particles $|\bm{Z}_n|=\sum_r Z^r_n$ 
       at generation $n$.
\item \emph{Relative growth rate}: the relative growth rate at generation $n$ 
       given by (for $n>1$)
       \[
        \mu(n)=\dfrac{|\bm{Z}_n|}{|\bm{Z}_{n-1}|}
       \]
       It is a multidimensional version of the Lotka-Nagaev estimator~\cite{L39,N67}, 
       which gives an empirical estimator of the \emph{malthusian parameter}.
\item \emph{Asymptotic distribution of classes}: the proportion of 
       particles in the $r$-th replicative class at generation $n$ given by
       \[
        u_r(n)=\dfrac{Z^r_n}{|\bm{Z}_n|} 
       \]
       The vector $\bm{u}(n)=\big(u_0(n),\ldots,u_R(n)\big)$ is called 
       \emph{asymptotic distribution of classes} 
       (or simply the \emph{class distribution}).
\item \emph{Average reproduction rate}: the average reproduction rate 
       (mean of the class distribution) at generation $n$ given by 
       \[
        \langle\varrho(n)\rangle=\sum_{r=0}^R r \, u_r(n)
       \]
       It can be shown that the average reproduction rate equals to the relative
       growth rate:
       \[
        \langle\varrho(n)\rangle=\mu(n) \qquad \text{for all}\quad n>1 
       \]
       (see Appendix~\ref{sec:APP1} for details).
\item \emph{Phenotypic diversity}: the variance (or standard deviation) of the  
       class distribution at generation $n$ given by  
       \[
        \sigma_{\varrho}^2(n)=\sum_{r=0}^R r^2 \, u_r(n) - \langle\varrho(n)\rangle^2
       \]
\item \emph{Phenotypic entropy}: the informational or Shannon entropy of the  
       class distribution at generation $n$ given by  
       \[
        h_{\varrho}(n)=-\sum_{r=0}^R u_r(n)\,\ln u_r(n) 
       \]
       Here we use the convention ``$0\ln 0 \equiv 0$''.
       This quantity behaves very much like the phenotypic diversity.
\item \emph{Normalized populational variance}: the normalized populational variance
       at generation $n$ given by
       \[
         \phi(n) = \sigma^2(n) - \sigma^2_{\varrho}(n)
       \]
       where $\sigma^2$ is the empirical estimator of the variance corresponding
       to the \emph{malthusian parameter} $\mu(n)$ (see Appendix~\ref{sec:APP1} 
       for details).
\end{itemize}

Strictly speaking, a surviving population described by branching process which does 
not becomes extinct grows indefinitely, at an exponential rate proportional to $\mu^n$.
Hence, in order to simulate a branching process it is necessary to impose a cut off 
on the progeny size, otherwise it would blow up the memory of the computer. 
This cut off is done by setting the \emph{maximum population size} $K$ which controls 
how much the population can grow unconstrained, acting in a similar fashion as the
\emph{carrying capacity} of the logistic growth~\cite{C03,L05}.
If the total number of particles that comprises the current generation is greater 
than the maximum population size $N$, a random sampling procedure is performed 
to choose $N$ particles to be used as parental particles for the next generation.
In particular, the progeny growth curve resembles a ``Logistic Growth Curve'' 
(see Figure~\ref{fig:fiebig}).

There are also some other additional settings that alter the way the program behaves. 
``Produce zero class particles'' allows to set if the particles of replicative 
capability $r=0$ will be considered in the calculations or not.
``Previous last generation/Do not preserve last generation'' 
allows to choose if the particles in previous generation will be carried over 
to current generation.
This was included in order to account for the possibility of a replication strategy 
that does not implement the disassemble of the parental particle.
In most cases the replication strategy used by RNA viruses implements the 
disassemble of the virus particle during the replication.
Retroviruses replication process is performed by the reverse transcriptase enzyme. 
The process of reverse transcription involves the synthesis of complementary DNA 
from the single-stranded RNA followed by the degradation of the intermediate RNA-DNA 
hybrid form.
The preservation of the parental generation in the model of viral evolution can allow
one or more particle to be preserved during several generations, in contrast with the
above-mentioned replication strategies of the RNA viruses.
The main routine of the program is given by the pseudo-code in Appendix~\ref{sec:APP3}. 

Finally, in order to discuss the simulations for the case when $b=0$ it is useful to 
introduce some conventions.
The \emph{instantaneous maximum replicative capability (at generation $n$)}, defined 
by $r_{\!*}(n)=\max\{r:Z_n^r\neq 0\}$, where $\bm{Z}_n=(Z_n^0,\ldots,Z_n^R)$ is the vector 
whose component $Z^r_n$ is the number of particles in the $r$-th replicative class at 
generation $n \geqslant 0$.
If the initial population $\bm{Z}_0=(Z_0^0,\ldots,Z_0^R)$ has $r_{\!*}(0)<R$ then all the 
quantities that depend on $R$ can must be calculated with $r_{\!*}(n)$ in the place of 
$R$, at the generation $n$.
Note that if $b=0$ then, for all purposes, $r_{\!*}=r_{\!*}(0)$ acts as the maximum 
replicative capability.
Even when $b \neq 0$, the parameter $r_{\!*}(n)$ acts as an ``instantaneous'' maximum 
replicative capability, which changes only when a particle in the highest replicative 
class $r_{\!*}(n)$ produces a progeny particle in the next replicative class, namely
$r_{\!*}(n+1)=r_{\!*}(n)+1$, that is retained in the population.

\section{Simulation Experiments}
\label{sec:CPM}

In this section we use some of the theoretical results from~\cite{ABCJ13b,ABCJ13a} 
(see also Appendix~\ref{sec:APP0}) to validate the \texttt{ENVELOPE} program at several 
levels of refinement. 
The validation is subdivided into several parts corresponding to distinct features of 
the model that are classified according to the possible regimes and phases of the time
evolution of a multitype branching process.
The subitems of the validation subsection can be read independently from each other. 
We also present new consequences of the combination of simulations with theoretical 
analysis and provide new perspectives on the role of fitness distributions and the 
variance of the branching process and on the mechanism of extinction, allowing us to 
propose a unifying principle underlying for the extinction of a virus population.

\subsection{Validation of the ENVELOPE Program}

\subsubsection{Transient Phase and Recovery Time}

\begin{figure}[ht] 
\begin{center}
 \includegraphics[scale=0.78]{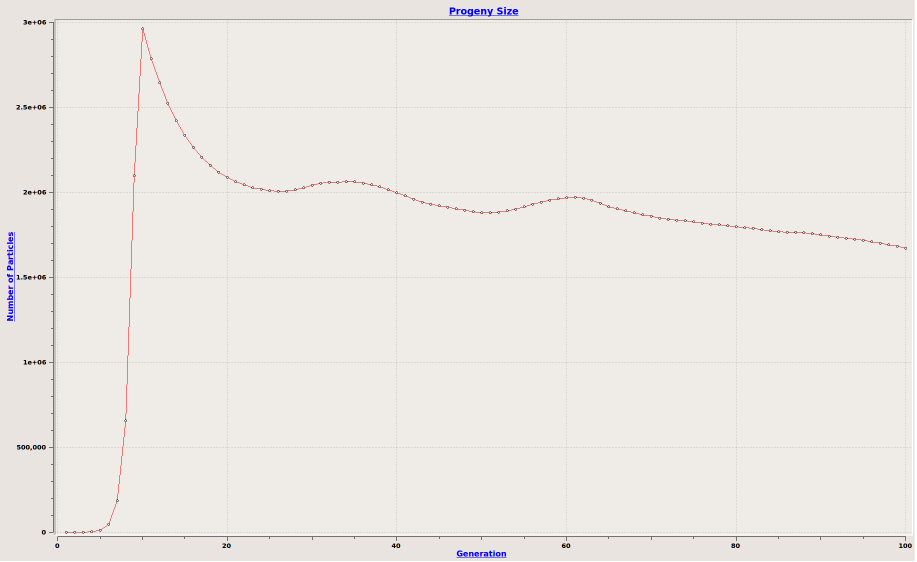}
 \caption{\label{fig:fiebig} Typical long term behavior of a population with two 
 distinct regimes. The first stage (from $n=0$ to $n$ around $10$--$20$) is the 
 transient stage and the second stage is the stationary stage.
 Parameters values: $b=0$; $d=0.75$; $Z_0^8=1$; $R=10$; $N=100$; $K=10^6$; 
 fitness distribution: Delta.} 
\end{center}
\end{figure}

A heterogeneous population replicating in a constant environment typically undergoes 
an initial period of high stochastic fluctuations in the relative frequency of each
variant, until it reaches a stationary regime where the relative frequencies become
constant.
This initial period, called \emph{transient phase}, is marked by the beginning
of the viral infection, after the bottleneck event when one or more particles are 
transmitted to a host organism and initiates the process of (re)establishment of 
the viral population in the new host.
The transient phase comprises the acute infection phase~\cite{FWR03,MBTGH10}, which 
is characterized by an initial exponential growth of the population, the attainment 
of the viremia peak, followed by a slower decrease towards a stabilization of the
population size (see Figure~\ref{fig:fiebig}).

In the phenotypic model the transient regime corresponds to the beginning of the time
evolution of the process. 
It is characterized, as noted before, by an instability of the relative frequencies 
of the replicative classes, an exponential growth of the progeny size, a decrease
of the average reproduction rate, and an increase of both the phenotypic diversity 
and the phenotypic entropy.

\begin{figure}[ht] 
\begin{center}
 \includegraphics[scale=0.78]{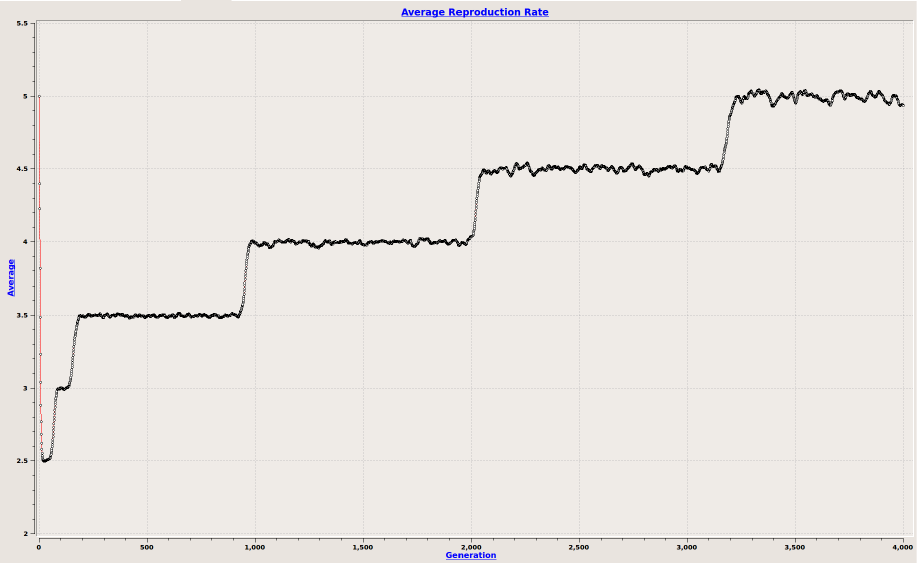}
  \caption{\label{fig:degrau} Average reproduction rate $=$ Relative growth rate.
  The ``jumps'' associated with the recovery time have heights about $0.5$.
  Parameters values: $b=0.000001$; $d=0.50$; $Z_0^5=1$; $R=10$; $N=4,000$; $K=10^6$;
  fitness distribution: Delta.} 
 \end{center}
\end{figure}

The expected time (as function of the number of generations) of the relaxation to-
wards an equilibrium after the bottleneck event, called \emph{recovery time}
(see the initial segment of the time series in Figure~\ref{fig:degrau}).

Let us assume, as usual, that the beneficial probability $b \approx 0$ and the founding 
population has $r_{\!*}(0)<R$.
Then one observes that the progeny size, the average reproduction rate, the phenotypic 
diversity and phenotypic diversity display a time series with several plateaus.
The ``length'' of each plateau is the number of generations that the population remains 
with the same value of $r_{\!*}(n)$ and the higher the plateau the longer, on average, 
is its length.
The presence of a jump indicates that a progeny particle form a parental particle in 
the replicative class $r_{\!*}(n)$ has undergone a beneficial effect, that is, the active 
maximum replicative capability increases by $1$ unit: $r_{\!*}(n+1)=r_{\!*}(n)+1$.
The occurrence of jumps can go on until $r_{\!*}(n)=R$.
Therefore, the ``length'' of each plateau represents the time, in number of generations, 
required for a beneficial effect to occur on a particle at the highest replicative class 
and be retained in the population.
The probability $\Prob\big(\text{jump in }r_{\!*}(n)\big)$ of occurrence of a jump event, 
when $r_{\!*}(n)<R$, may be estimated using equation~\eqref{eq:REIGENVECPERT} of
Appendix~\ref{sec:APP0} as
\[
 \Prob\big(\text{jump in }r_{\!*}(n)\big) \approx b \, u_{r_{\!*}(n)} 
 \approx b \, (1-d)^{r_{\!*}(n)}
\]
where $u_{r_{\!*}(n)}$ is the proportion of particles in the $r_{\!*}(n)$-th replicative 
class at generation $n$, which is the instantaneous maximum replicative capability 
at time $n$.
Notice that, as $r_{\!*}(n)$ increases, $u_{r_{\!*}(n)}$ decreases monotonically and 
therefore, $\Prob\big(\text{jump in }r_{\!*}(n)\big)\to 0$ when $r_{\!*}(n)\to\infty$.
This result highlights the asymmetry between the contributions of the beneficial
probability versus the deleterious probability to the recovery time.

The ``height'' of a jump in the average reproduction rate time series is independent
of the plateau where the jump occurs.
In order to estimate the ``height'', consider two consecutive levels on the time series 
of the average reproduction rate, the first ``height'' $\mu(n_1)$ measured at generation 
$n_1$ and the second ``height'' $\mu(n_2)$ measured at generation $n_2$, with $n_1<n_2$ 
not necessarily consecutive, such that $r_{\!*}(n_2)=r_{\!*}(n_1)+1$ and $\mu$ is 
approximately constant around $n_1$ and $n_2$.
Thus, the difference $\mu(n_2)-\mu(n_1)$ gives an estimate of the height of the jump 
between two consecutive plateaus.
When $b \approx 0$, equation~\eqref{eq:PERTEXPAN} of Appendix~\ref{sec:APP0} implies 
that $\mu(n) \approx r_{\!*}(n) (1-d)$ and hence
\[
 \mu(n_2)-\mu(n_1) \approx \big(r_{\!*}(n_2)-r_{\!*}(n_1)\big) (1-d)
 \approx 1-d \,.
\]
For instance, in Figure~\ref{fig:degrau} it can be readily seen that the height of the 
jumps is about $0.5$ and, in fact, $d=0.50$, $b=0.000001$ and hence $1-d=0.4999999$.

\subsubsection{Stationary Regime}

The advanced stage of the infection, also called chronic infection
phase~\cite{FWR03,MBTGH10}, is comprised by the stationary regime where the viral
population has recovered its phenotypic (and genotypic) diversity and becomes better
adapted to the new host environment by exhibiting rather stable relative frequencies 
of almost all variants.

In the phenotypic model the stationary regime corresponds to the asymptotic behavior 
of a super-critical branching process ($\mu>1$).
But, as mentioned before, a surviving population described by super-critical branching
process is never stationary (in the strict sense) and therefore this correspondence is 
not straightforward.

The \emph{normalized process} $\bm{W}_n=\bm{Z}_n/\mu^n$ is stationary and, when 
$n\to\infty$, the random variable $Z^r_n/|\bm{Z}_n|$ converges to the asymptotic 
relative frequency $u_r$ of $r$-th replicative class.
Consequently, the average reproduction rate $\langle\varrho(n)\rangle=\mu(n)$, 
the phenotypic diversity $\sigma^2_{\varrho}(n)$ and the phenotypic entropy 
$h_{\varrho}(n)$ remain essentially constant in time.
Moreover, the \emph{maximum population size} cut off $K$ ensures that the total 
progeny size remains constant in time with expected value 
$\langle |\bm{Z}_n|\rangle \approx \mu(n)\,K$.

During the stationary regime, the stability of the relative frequency of each class 
is maintained by a steady ``flow of particles'' from a replicative class to its 
adjacent classes, due to the deleterious probability $d$ and the beneficial 
probability $b$. 
The probability $c$ contributes maintenance of a constant proportion of particles 
in each replicative class.
When the beneficial probability $b \neq 0$ the asymptotic distribution of classes 
$u_r$ is independent of the configuration of the founding population and, when
$n$ is large enough, $r_{\!*}(n)=R$. 

\begin{figure}[ht]
\begin{center}
 \includegraphics[scale=0.78]{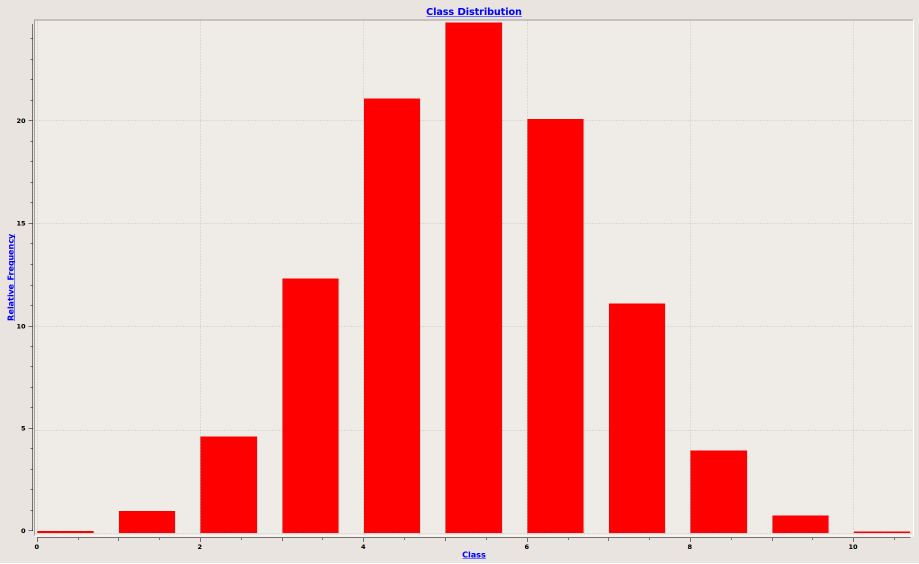}
 \caption{\label{fig:equilibrio} Histogram of the replicative classes. 
 Parameter values: $b=0.000001$; $d=0.50$; $Z_0^5=1$; $R=10$; $N=4,000$; $K=10^6$;
 fitness distribution: Delta.}
\end{center}
\end{figure}

More importantly, when $b \approx 0$, the replicative classes that are most 
representative in the population are the classes near the mode of the distribution
of classes $u_r$, also known as ``most probable replicative capability''.
The mode of $u_r=\mathrm{binom}(r;R,1-d)$ is given by $m(u_r)=\lfloor (R+1)(1-d)\rfloor$, 
except when $(R+1)(1-d)$ happens to be an integer, then the two replicative classes 
corresponding to $(R+1)(1-d)-1$ and $(R+1)(1-d)$ are equally ``most probable''
(see~\cite{F68}, here, $\lfloor x\rfloor$ denotes the greatest integer less than $x$).
When $(1-d) \approx 1/2$ the mode is close to the average reproduction rate 
$\mu(n)=\langle\varrho(n)\rangle$ (see Figure~\ref{fig:equilibrio}).

\subsubsection{Threshold of Extinction}
 
The threshold of extinction takes place when the deleterious rate is sufficiently high
that it prevents the viral population of reaching the stationary regime but not high
enough to induce the extinction of the population in the short run.
Therefore, any small increase in the deleterious rate can push the population toward
extinction, while any small decrement can allow the population to reach the stationary
regime. 
 
In the phenotypic model, the threshold of extinction corresponds to a critical branching
process ($\mu=1$) and is characterized by instability of the relative frequencies of 
the replicative classes, the average replicative rate and the phenotypic diversity.
The instability observed represents the impossibility of the viral population to 
preserve, due to the deleterious effects, particles with high replicative capability.
The occurrence of an eventual extinction of the population is almost certain, although 
the time of occurrence of the extinction may be arbitrarily long if the initial
population is sufficiently large.
In other words, the threshold of extinction looks like an infinite transient phase and
is the borderline between the stationary regime, where the transient phase ends at an
stationary equilibrium, and the extinction in finite time.

\begin{figure}[ht] 
\begin{center}
 \includegraphics[scale=0.78]{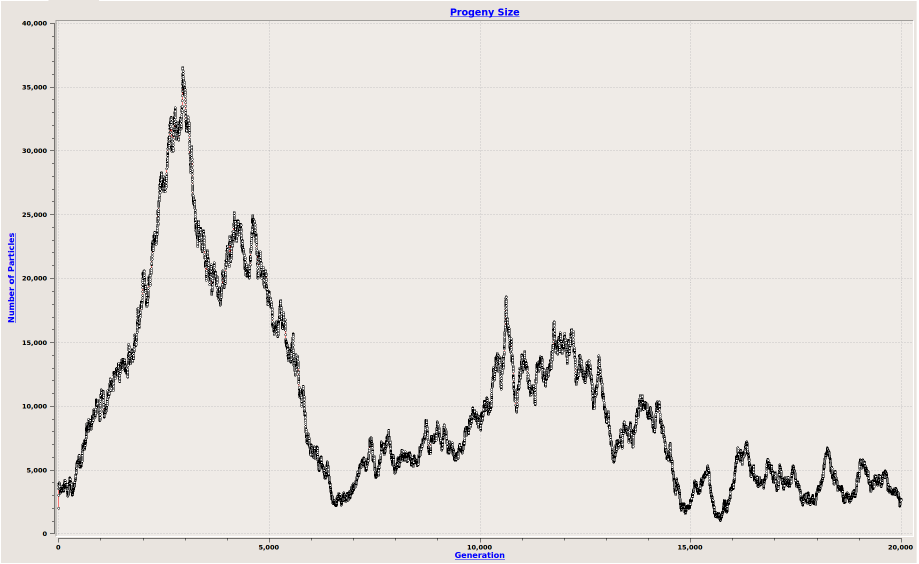}
 \caption{\label{fig:critico} Progeny size of a population at the extinction threshold. 
   Parameter values: $b=0$; $d=0.50$; $Z_0^2=1,000$; $R=2$; $N=20,000$; $K=10^6$;
   fitness distribution: Delta.} 
\end{center}
\end{figure}

Setting the parameters of the phenotypic model in order to obtain a critical branching 
process is a matter of ``fine tuning'', since it requires that the probabilities $d$, 
$b$ and the maximum replicative capability $R$ satisfy the algebraic equation 
$\mu(b,d;R)=1$ -- which is a non-generic condition (see Figure~\ref{fig:critico}).

When $b=0$, the critical deleterious probability is $d_{c}=1-1/R$, for each fixed $R$.
When $b \neq 0$ one may consider, for fixed $R$, the corresponding critical probability 
$d_c(b)$ as given implicitly by the equation $\mu(b,d_c(b),R)=1$ and the condition 
$\mu(0,d_c(0),R)=1$, with $d_{c}(0)=1-1/R$ (see Figure~\ref{fig:simplex} of Appendix~\ref{sec:APP0}).
Since $b$ and $d$ are constrained to satisfy $b+d \leqslant 1$, there is a maximum
value of $b$ such that $b+d_c(b)=1$, for each fixed $R$.
Denote this maximum value by $b^\star(R)$.
The number $b^\star(R)$ is the maximum beneficial probability such that the phenotypic 
model has three distinct regimes.
In other words, if $b>b^\star(R)$ then $d<d_c(b)$ and the process never becomes extinct.  
In the parameter space of the phenotypic model, the critical probabilities $d_c(b^\star)$ 
at $b^\star$ are given as the intersection of the boundary line $b+d=1$ with the 
critical curves $\mu(d,b,R)=1$ for each fixed $R$.

Using the expressions for the malthusian parameter obtained in~\cite{ABCJ13a}, 
it is easy to show that the following approximations hold (when $R\to\infty$)
\begin{equation} \label{eq:APPFOR}
\begin{split}
 b^\star(R) & \approx \dfrac{1}{R} - \dfrac{1}{R}\,\dfrac{(R-1)^2}{1+(R-1)^2} \\
 d_c(b^\star(R)) & \approx d_c(0) + \dfrac{1}{R}\,\dfrac{(R-1)^2}{1+(R-1)^2} \,.
\end{split} 
\end{equation}
Here, one uses that $b^\star+d_c(b^\star)=1$ and $d_c(0)=1-1/R$.
Comparison of critical deleterious probability given by equations~\eqref{eq:APPFOR} 
with the correct values obtained by numerical computation using the mean matrix, 
shown in Table~\ref{tab:TAB1}, indicate that the asymptotic expressions converge
to the real values when $R\to\infty$.

\renewcommand{\arraystretch}{1.5}

\begin{table}[ht]
\begin{center}
\begin{tabular}{c@{\qquad}c@{\qquad}c@{\qquad}c@{\qquad}c} \hline
 $R$ & $d_c(0)$ & $d_c(b^\star)$ & 
 $\tilde{d}_c(b^\star)$ & $|d_c(b^\star)-\tilde{d}_c(b^\star)|$ \\ \hline
  2 & 0.50 & 0.707 & 0.750 & 0.043 \\
  3 & 0.66 & 0.895 & 0.933 & 0.038 \\
  4 & 0.75 & 0.951 & 0.975 & 0.024 \\
  5 & 0.80 & 0.972 & 0.988 & 0.016 \\
  6 & 0.83 & 0.982 & 0.993 & 0.011 \\ \hline
\end{tabular}
\end{center}
\caption{\label{tab:TAB1} Critical deleterious probabilities $d_c(0)$ and $d_c(b^\star)$. 
         The real values of $d_c(b^\star)$ were obtained by numerical computation using 
         the mean matrix and the values denoted by $\tilde{d}_c(b^\star)$ were obtained 
         using equations~\eqref{eq:APPFOR}.}
\end{table}

\renewcommand{\arraystretch}{1}

\subsubsection{Extinction by Lethal Mutagenesis}
 
The process of extinction of the viral population induced by increase of the 
deleterious rate is called \emph{lethal mutagenesis}~\cite{BSW07}.
In the phenotypic model, the lethal mutagenesis corresponds to a sub-critical 
branching process ($\mu<1$). 
It is characterized by continuous decrease of the average replicative rate 
and by increase of the phenotypic diversity followed by a sudden decrease in 
the subsequent generations.

The progeny size and the phenotypic diversity increase during the first generations
because the founding population still has reasonable replicative capability.
However, increasing the size of the founding population does not prevent extinction, 
it only increases the time required for the extinction to occur.  
Increasing the deleterious probability $d$ decreases the time required for 
extinction and increasing beneficial probability $b$ can prevent extinction.

Note that when $b=0$ the population cannot achieve a replicative capability higher 
than the one present in the founding population.  
In this case, a population transmitted to a new host organism via a bottleneck event 
will have maximum replicative capability less or equal to the maximum replicative 
capability of the original population.

\begin{figure}[ht] 
\begin{center}
 \includegraphics[scale=0.78]{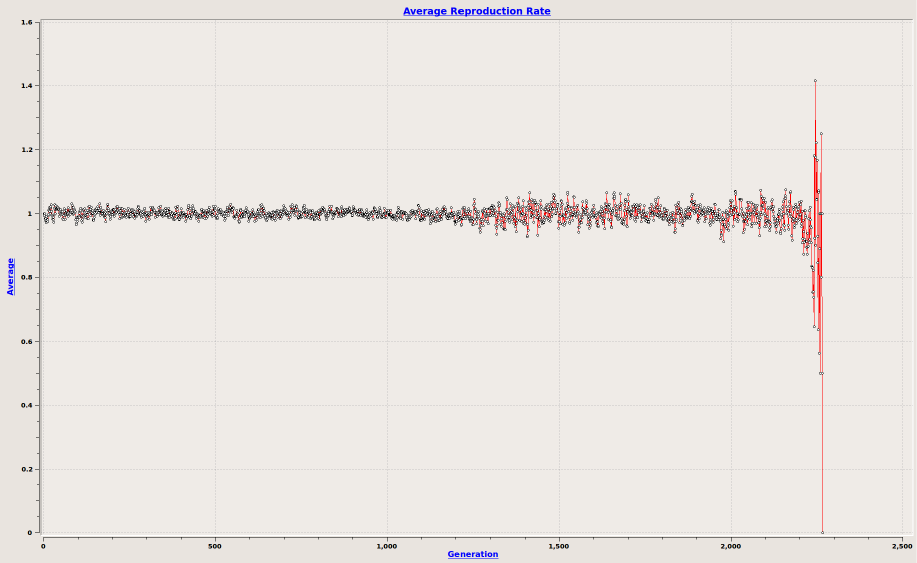}
 \caption{\label{fig:lethal} Lethal mutagenesis and the path to extinction. 
   Parameter values: $b=0$; $d=0.501$; $R=2$; $N=2,500$; $K=10^6$;
   $\bm{Z}_0=(1000,2000,1000)$; fitness distribution: Delta.} 
\end{center}
\end{figure}

Interesting enough, there is a signature of the extinction process which may be 
directly observed in the behavior of the average reproduction rate curve $\mu(n)$.
It is marked by an explosive growth in the variation of $\mu(n)$ as $n$ approaches 
the extinction time $n^*$ (see Figure~\ref{fig:lethal}).

The phenomenon of explosive growth near the extinction event may be detected by the 
\emph{oscillation} of $\mu(n)$ in an interval ending at the last non-zero generation:
\[
 \mathrm{osc}(\mu)=\max_{n<n^*} \mu(n) \; - \min_{n<n^*} \mu(n) \,.
\]
Even when the process is slightly super-critical, it is expected that the oscillation
of $\mu(n)$ remains very small, with $\mathrm{osc}(\mu)\sim 10^{-3}$ for all $n$. 
On the other hand, when a slightly sub-critical process is approaching the 
extinction time $n^*$ one typically observes $\mathrm{osc}(\mu) \sim 10^{-1}$.

The \emph{expected time to extinction} $\langle T_{\mathrm{ext}}\rangle$ of a branching
process was determined in~\cite{JKS07}: if $\mu\leqslant 1$ then
\[
 \langle T_{\mathrm{ext}}\rangle = \dfrac{\ln Z_0^{r_{\!*}}+\kappa}{-\ln\mu}
\]
where $\kappa>0$ depends only on the parameters of the model 
(not on the initial population).
It is easy to show that at the critical value of the malthusian parameter ($\mu=1$)
equilibrium is never reached.
A scaling exponent characterizing the behavior of expected time to extinction in
a neighborhood of the critical value of the malthusian parameter can be obtained by
considering the first order expansion of $\langle T_{\mathrm{ext}}\rangle$ about $1$:
\[
 \langle T_{\mathrm{ext}} \rangle \approx |\mu-1|^{-1} \,.
\]
When $b=0$ one may write $\langle T_{\mathrm{ext}} \rangle$ as a function of the
deleterious probability and the critical deleterious probability $d_c=1-1/R$ as
\[
 \langle T_{\mathrm{ext}} \rangle \approx \frac{1}{R}|d-d_c|^{-1}
\]
since $|\mu-1|=R|d-d_c|$.
This ``scaling law'' is formally identical to the one obtained in~\cite{GD15} for the 
error threshold of the deterministic quasispecies model as a function of the mutation 
rate.

\subsection{Populational Variance and the Role of Fitness Distributions}

All properties of the phenotypic model that have been discussed so far are related 
to the mean matrix of the model, that is, they depend only on the first moments of 
the branching process and may be called ``first order properties''.
In particular, they are independent of the choice of the family of fitness 
distributions.
If we want to see how the fitness distributions influence the evolution of the 
population we must to look at a ``second order property'', which is expected to 
depend on the second moments of the fitness distributions (see Table~\ref{tab:TAB3}).

\renewcommand{\arraystretch}{1.5}
\begin{table}[ht]
\begin{center}
\begin{tabular}{c@{\qquad}c@{\qquad}c} \hline     
 Distribution & $\sigma^2_r$ & $\phi$ \\ \hline
 Deterministic & $0$ & $0$ \\
 Poisson & $r$ & $\langle\varrho\rangle$ \\
 Geometric & $r(r+1)$ & 
 $\langle\varrho\rangle\big(\langle\varrho\rangle+1\big)+\sigma^2_{\varrho}$ \\
 $\tfrac{1}{2}$-Binomial & $r/2$ & $\langle\varrho\rangle/2$\\
 Power law & $+\infty$ & $+\infty$ \\
\hline
\end{tabular}
\end{center}
\caption{\label{tab:TAB4} Location-scale families of fitness distribution of the
         \texttt{ENVELOPE} program, their variance and the corresponding normalized 
         populational variances.}
\end{table}
\renewcommand{\arraystretch}{1}

The simplest property of second order is given by the population variance $\sigma^2$
associated with the malthusian parameter $\mu$ (namely, the relative growth rate). 
Furthermore, the difference between the populational variance and the (squared) 
phenotypic diversity, called \emph{normalized populational variance} and denoted 
by $\phi$ is a very interesting quantity to be measured, since it satisfies 
\begin{equation} \label{eq:PHISIGMA}
 \phi = \sigma^2 - \sigma^2_{\varrho} = \sum_{r=0}^R \sigma^2_{r}\,u_r 
\end{equation}
In other words, $\phi$ is a weighted average of the variances $\sigma^2_r$ of the 
fitness distributions. 
See Appendix~\ref{sec:APP1} for the precise definition of $\sigma^2$ and the proof 
of the second equality in equation~\eqref{eq:PHISIGMA}.

Given a location-scale family of fitness distributions $t_r$ such that $\sigma_{r}^2$ 
is at most a quadratic polynomial on $r$, eqaution~\eqref{eq:PHISIGMA} allows one to 
write the corresponding normalized population variance $\phi$ in terms of the average
reproduction rate $\langle\varrho\rangle$ and the phenotypic diversity $\sigma_{\varrho}^2$.
Hence, $\phi$ can be exactly computed for all location-scale families of distributions 
used in the \texttt{ENVELOPE} program (see Table~\ref{tab:TAB4}).

It is important to stress that unlike the malthusian parameter, the normalized 
populational variance does depend on the choice of the family of fitness distributions.
Recall that the malthusian parameter depends only on the mean matrix, which depends 
on the fitness distributions $t_r$ only through its expectation values.
Since we have imposed the same normalization condition that the expectation value of
$t_r$ is $r$ for all families of fitness distributions, it follows that the mean matrix, 
and hence the malthusian parameter, does not depend on the family of fitness 
distributions.
On the other hand, the variances of different families of fitness distributions are not 
necessarily the same.
For instance, if $t_r$ is the family of Poisson distributions then $\sigma^2_{r}=r$
and thus $\phi=\mu$.

Assume that $b=0$ (then $c=1-d$).
From the expression of the asymptotic distribution of classes~\eqref{eq:REIGENVEC} 
one obtains: $\langle\varrho\rangle=\mu=R(1-d)$ and $\sigma_{\varrho}^2=Rd(1-d)$.
Moreover, when $b \neq 0$ is sufficiently small, formula~\ref{eq:REIGENVECPERT} 
ensures that $\langle\varrho\rangle$ and $\sigma_{\varrho}^2$ are approximated 
by the corresponding values for $b=0$ and the same holds for $\phi$. 

\begin{figure}[ht] 
\begin{center}
 \includegraphics[scale=0.78]{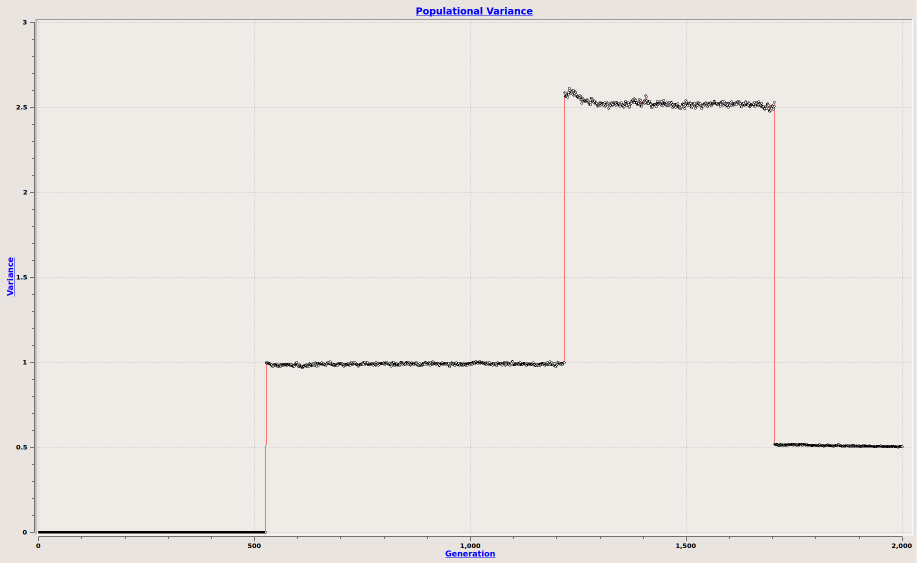}
 \caption{\label{fig:variance} Normalized population variance $\phi$, 
          with $\langle\varrho\rangle=1$ and $\sigma^2_{\varrho}=0.5$. 
          Parameter values: $b=0$; $d=0.50$; $R=2$; $N=2,000$; $K=10^6$;
          $Z_0^2=10,000$; fitness distributions: Delta ($\phi=0$), 
          Poisson ($\phi=1$), Geometric ($\phi=2.5$), Binomial ($\phi=0.5$).} 
\end{center}
\end{figure}

For instance, in Figure~\ref{fig:variance} we show the graph of the normalized 
population variance $\phi(n)$, at generation $n$, from a simulation in which we 
switched among the four families of fitness distributions with finite variance
using the ``Video'' function of the \texttt{ENVELOPE} program to pause the 
simulation and change the type of fitness distribution.

\begin{figure}[ht] 
\begin{center}
 \includegraphics[scale=0.78]{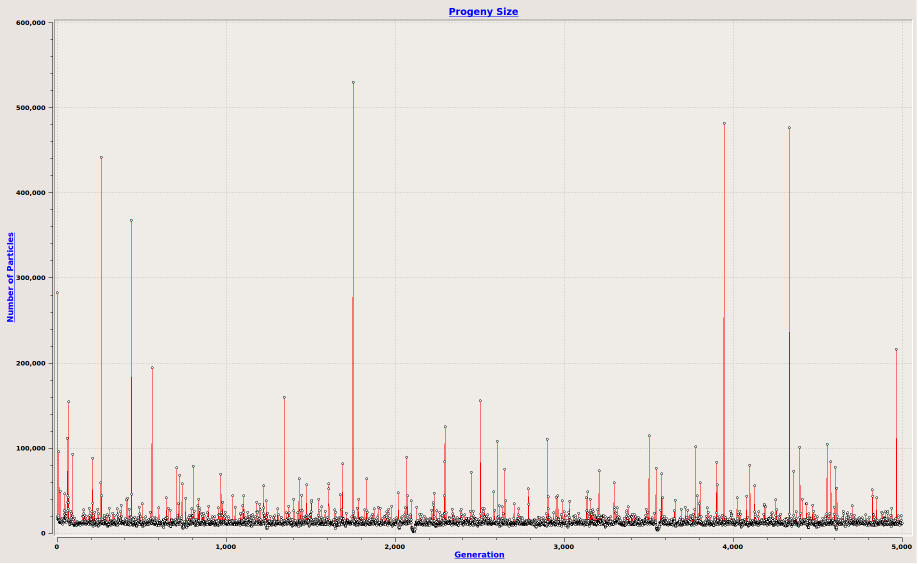}
 \caption{\label{fig:blips} Power law induced bursts (viral blips), 
   with time-average progeny size around $14,000$ viral particles,
   going up to $400,000$ - $500,000$ particles during an intense burst.
   Parameter values: $b=0$; $d=0.30$; $R=2$; $N=5,000$; $K=10^4$;
   $Z_0^2=10,000$; fitness distribution: Power law.} 
\end{center}
\end{figure}

Finally, it is worth to remark that the impact of the power law family of fitness 
distribution on the evolution of the population is very distinct from the other 
families, because, unlike the other fitness distributions, it has infinite variance.
One of the consequences of this property is the appearance of intense bursts of 
progeny production clearly seen on the times series of progeny size and the average 
reproduction rate (see Figure~\ref{fig:blips}).
The instability caused by unbounded fluctuations coupled with the finite population 
size effect (even for large $K$) is responsible for the generation of a train 
of sparse and intense bursts of progeny production.
On the other hand, this instability coupled with finiteness effect may also 
provoke sudden drops on the progeny size driving the population to a premature 
extinction, even if the malthusian parameter is above $1$.
Because of these extreme phenomena one would be led to believe that the phenotypic 
model with the power law family of fitness distributions is an exception to the 
general result: any property derived from the mean matrix is independent of the
fitness distribution.
It is not the case.
In fact, if one considers the \emph{time-average} of any quantity that is 
time-dependent over a time interval $[n_0,N]$ during the stationary regime, 
let's say
\[
 \bar{\mu}(N) = \frac{1}{N} \sum_{n=n_0}^N \mu(n) \,,
\]
then it is expected that $\bar{\mu}(N)$ becomes very close to the asymptotic 
value of the relative growth rate $\mu$ when $N$ is sufficiently large.

For instance, in Figure~\ref{fig:blips} the time-average of the progeny size 
$\bar{\mu}(N)$ is around $14,000$, while the expected progeny size for the model 
is $\mu K=R(1-d)K=2 \times 0.7 \times 10^4=1.4 \times 10^4=14,000$, in full 
agreement with the general theory.

\subsection{Finite Population Size and Mutational Meltdown}
\label{subsec:FPSMM}

Recently, Matuszewski \etal~\cite{MOBJ17} reviewed the literature about theories and
models describing the extinction of populations owing to the excessive accumulation 
of deleterious mutations or effects and distinguished two apparently distinct lines of 
research, represented by the lethal mutagenesis models~\cite{BSW07} and the mutational 
meltdown models~\cite{LG90} which, nonetheless, display a considerable amount of 
similarity.

Indeed, as shown in~\cite{BSW07,ABCJ13b,ABCJ13a}, lethal mutagenesis is independent of 
population size, hence it is fundamentally a deterministic process that operates even 
on very large populations.
Although the outcome of lethal mutagenesis is deterministic, other aspects of the
population dynamics (such as extinction time, individual trajectories of progeny 
size, etc.) are not.
On the other hand, the mutational meltdown generally works within the context of
``small'' population sizes in which stochastic effects caused by random drift play 
an important role.

We believe that the approach presented here may help shed some light on this issue.
There is one ingredient in the mutational meltdown theory that is absent in the
lethal mutagenesis theory: the carrying capacity.
This is true even for models with finite population, such as~\cite{DSS85} and
the phenotypic model, in their theoretical formulations as branching process.
However, as seen before, the computational implementation of the phenotypic model 
required the introduction a cut off $K$ in order to bound the growth of the population.
If the cut off is taken as basic constituent of the phenotypic model, and not merely a 
convenient device, then it can play a role similar to a carrying capacity and the model 
may no longer be considered a ``pure'' branching process, but a \emph{self-regulating}
branching process~\cite{MS12,MSR13}.

In a self-regulating branching process not all the offspring produced in a given 
generation will to produce offspring in the next generation and hence, it is necessary 
to introduce a \emph{survival probability distribution} $S(n|T_n)$, to stochastically 
regulate the survival of offspring at any generation $n$ as a function of the total 
population size $T_n=|\bm{Z}_n|$.
The motivation behind this definition is the following: if the population size at 
a generation $n$ exceeds the carrying capacity of the environment then, due to 
competition for resources, it is less likely that an offspring produced in that
generation will survive to produce offspring at generation $n+1$.

Let $S(n|T_n)$ denote the conditional probability that any offspring produced
at generation $n$ survives to produce offspring at generation $n+1$, given that
the population has $T_n$ individuals at generation $n$.
If we define the conditional probability $S$ as
\[
 S(n|T_n)=\left\{\begin{array}{l@{\quad\text{if}\quad}l}
        K / T_n & T_n > K \\
        1 & T_n \leqslant K
 \end{array}\right.
\]
then the phenotypic model becomes a self-regulating process with carrying capacity $K$.
Moreover, when $K\to\infty$ the self-regulating process reduces to a ``pure'' 
branching process.

If $K$ is not large enough then a kind of \emph{random drift} effect due to finite 
population size may take place, which happens when the fittest replicative classes 
are lost by pure chance, since its frequency is typically very low (they are the 
lesser represented replicative class in the population).
If the loss of the fittest replicative class occurs a sufficient number of times then
the population will undergo extinction.
Note that this may happen even when the process is super-critical, namely, it is far
from the extinction threshold.
This is not a contradiction with the definition of extinction probability, since a
super-critical process still has a positive probability to become extinct
(see Appendix~\ref{sec:APP1}).

Now suppose that $b=0$, the initial population has active maximum replicative capability 
$r_{\!*}(0)$ and the carrying capacity $K$ is sufficiently small (we shall give an 
estimate of $K$ in a moment).
Then, as mentioned before, the value $r_{\!*}=r_{\!*}(0)$ acts as the maximum replicative 
capability for that population.
Moreover, if the highest replicative class $r_{\!*}$ is lost by chance, that is, if 
$r_{\!*}(n+1)=r_{\!*}(n)-1$, then it can not be recovered anymore and hence, from that
time on the maximum replicative capability for that population has dropped by $1$ unit.
This may be seen as a manifestation of the ``Muller's ratchet'', since the population 
has accumulated a deleterious effect in an irreversible manner.

\begin{figure}[ht] 
\begin{center}
 \includegraphics[scale=0.78]{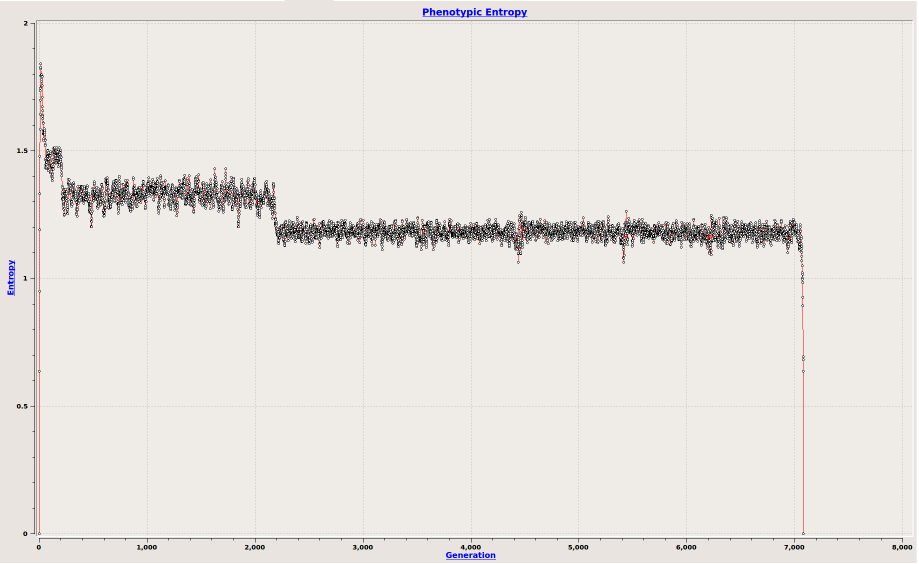}
 \caption{\label{fig:ratchet} Extinction by mutational meltdown. 
          Phenotypic Entropy time series.
          At the beginning the branching process is super-critical with $\mu=3.41$.
          Parameter values: $b=0$; $d=0.659$; $R=10$; $N=10,000$; $K=2,000$;
          $Z_0^{10}=6,000$; fitness distributions: Delta.} 
\end{center}
\end{figure}

For sake of concreteness, let us assume that $r_{\!*}=R$ and $d$ are such that 
$(R-1)(1-d)<1$, but $R(1-d)>1$.
Then, at the beginning of the process, the malthusian parameter is $\mu=R(1-d)>1$ and
the process is super-critical.
However, if at some generation $n$, the $R$-th replicative class is lost by chance, 
then $R$ drops by $1$ and $\mu=(R-1)(1-d)<1$, so the process becomes sub-critical and
the population becomes extinct very quickly.
In this case, the frequency of the $R$-th replicative class is $(1-d)^{R}$ and 
fraction of particles that are purged, at each generation, is $R(1-d)-1$, hence
the fraction of particles that are left in the $R$-th replicative class, at
each generation, is $\nu_{R}=2(1-d)^{R}-R(1-d)^{R+1}$.
If $K\approx 1/\nu_{R}$ then there will be, on average, $1$ particle of class $R$ 
per generation -- it is very unlikely that this replicative class will be retained
for a long period of time.
Therefore, in order to avoid the random drift effect $K$ should be at least of the 
order of $10 \times R(1-d)/\nu_{R}$, or higher.
At each ``click of the ratchet'' the fittest replicative class is lost and there is a 
drop in the malthusian parameter by $(1-d)$, until $r_{\!*}(1-d)$ becomes less than $1$, 
where $r_{\!*}$ is the maximum replicative capability at the current generation.
This drop occurs in the phenotypic diversity and the phenotypic entropy, as well 
(see Figure~\ref{fig:ratchet}).

If one writes the usual condition for occurrence of extinction $r_{\!*}(1-d)<1$ as
\[
(1-d) < 1/r_{\!*}
\]
then this is an exact phenotypic analogue of the mutational meltdown extinction 
criterion.
Indeed, $r_{\!*}$ is the phenotypic analogue of \emph{absolute growth rate} of the 
population at time $n$ and $c=(1-d)$, the probability of occurrence of a neutral 
fitness effect per individual particle, is the phenotypic analogue of the 
\emph{mean viability} (compare with the equations in~\cite{LBBG93,LG90}).

\section{Conclusion and Outlook}
\label{sec:CD}

In this paper we have exhaustively explored a model for the evolution of RNA virus, 
which was formulated as a multivariate branching process, called \emph{phenotypic model}.
The theory of branching processes provides a suitable framework endowed with concepts 
and analytic tools allowing for the investigation of evolutionary aspects of RNA viruses
propagating along different adaptive landscapes.  

One of the greatest virtues of the phenotypic model is its simplicity.
Since the model has essentially only $3$ parameters it is possible to analytically 
compute the spectrum of its mean matrix and, applying the classification of multitype 
branching processes, obtain a complete qualitative description its ``generic behaviors'', 
that is, the most likely outcomes of the model's asymptotic dynamics.

The maximum replication capacity $R$ and the probabilities of occurrence of deleterious
effects $d$ entirely determine whether a viral population becomes extinct infinite time
or not. 
On the other hand, the third parameter, the probability of occurrence of beneficial
effects $b$, plays a distinct role from the other two probabilities, functioning as 
a threshold parameter which determines if the model posses the three typical regimes 
of a branching process or just one regime (super-critical).

The model provides several statistical measures, such as average growth rate, 
phenotypic diversity, phenotypic entropy and population variance, that allows one 
to asses the stochastic dynamics of a viral population.
The dynamics of the associated deterministic quasispecies model is given by a mean 
field limit where the mean matrix completely determines the dynamics (see
Appendix~\ref{sec:APP1/2}).
Hence it is possible to establish a relation between statistical measures mentioned 
above and the fundamental macroscopic parameters that characterize the evolutionary
dynamics of a quasispecies.
In particular, the models of~\cite{SS82,SS1988} could be used as a representation 
of the evolution of mean values obtained from the mean matrix of a branching process.
By extending the scope of the model to \emph{age dependent branching processes}~\cite{AN72}
could allow the incorporation other statistical measures, such as evolutionary 
entropy~\cite{D05,D2013}.
This quantity could provide a more precise understanding of viral diversity 
given the fact that population sizes of viral population are finite.

Despite its conceptual appeal, the phenotypic model has some important drawbacks. 
The first limitation is the lack of feedback from the host organism on the virus
population, since the probabilities of fitness effects  are independent of time. 
This shortcoming is partially handled in the \texttt{ENVELOPE} program by the ``Video''
function, which allows one to pause the simulation and change the probabilities and
emulate the host's ``response'' against  the virus. 
The second limitation is the lack of the phenotype-to-genotype map, i.e, the
relationship between genotypic and phenotypic change.
The motivation to use a phenotypic approach was to avoid the severe difficulties 
in modeling this kind of mapping~\cite{A91,FZOW17}. 

Even though there is no phenotype-to-genotype map, it still is possible to draw
some consequences about mutation rates from a purely phenotypic model~\cite{ABCJ13b}.
For example, under the assumption that the mutation rate $U$ is sufficiently high 
(between $0.1$ and $1$), the probability that a spontaneous mutation produces 
a deleterious effect may be estimated as follows: if we assume that the number 
of mutations in a genome follows a Poisson distribution, 
then $d_{\mathrm{sp}} \approx 1-\ee^{-f_d}$, where $f_d$ is 
the probability that a spontaneous mutation has a deleterious effect~\cite{KM66}. 
Values of $f_d$ have been measured \emph{in vitro} for a few viruses and are shown 
in Table~\ref{tab:COMP}, along with the respective mutation rates $U$.
Now, $d_{\mathrm{sp}}$ provides a lower bound for the deleterious probability and since
the value $d_{\mathrm{sp}} \approx 1/2$ seems to be typical for RNA viruses, the interval 
$1/2<d<d_c=1-1/R$ is more likely to be the range of the parameter $d$.
Moreover, it is easy to see that the phenotypic diversity and the phenotypic entropy 
are maximal when $d$ is near $1/2$, for any value of $R$~\cite{ABCJ13b}.
One could speculate that this is a universal property for RNA viruses that replicate
under high mutational rates associated with a maximization principle that seeks to
improve the chances of survival~\cite{D12}.

\renewcommand{\arraystretch}{1.5}
\begin{table}[ht]
\begin{center}
\begin{tabular}{c@{\qquad}c@{\qquad}c@{\qquad}c@{\qquad}c@{\qquad}c} \hline
 Virus & Group & $f_d$ & $d_{\mathrm{sp}}$ & $U$ & REFS. \\ \hline
 VSV & ($-$)ssRNA (V) & $0.69$ & $0.50$ & $0.1$ & \cite{SME04,FMS05} \\
 Q$\beta$ & ($+$)ssRNA (IV) & $0.74$ & $0.52$ & $0.6$ & \cite{CCS09,BCDS13} \\ 
 TEV & ($+$)ssRNA (IV) & $0.77$ & $0.53$ & $0.5$ & \cite{CIE07,TE10} \\
 $\Phi$6 & dsRNA (III) & $0.42$ & $0.34$ & $0.03$ & \cite{BC04,BGSS07} \\ 
 $\Phi$X174 & ssDNA (II) & $0.77$ & $0.53$ & $0.003$ & \cite{CCS09,CDS09} \\
 F$1$ & ssDNA (II) & $0.65$ & $0.47$ & $0.004$ & \cite{PDC10,D12} \\ \hline
\end{tabular}
\end{center}
\caption{\label{tab:COMP} Measured values of $f_d$, the corresponding deleterious 
 probability $d_{\mathrm{sp}}=1-\ee^{-f_d}$ and genome-wide spontaneous mutation rate $U$.}
\end{table}
\renewcommand{\arraystretch}{1}

The first main result of this paper is concerns the role of the fitness distributions.
The fitness distributions of the phenotypic model were motivated by the results 
and observations of~\cite{ZYY09} on the distribution of single cell progeny sizes 
of RNA viruses. 
In~\cite{ZYY09} the authors demonstrated that even in a well controlled experiment, 
using the same viral isolate, same infection parameters and clonally expanded target 
cells, progeny sizes can vary substantially. 
The variance on progeny sizes in such uniform environment indicates that RNA viruses
replication bears in some way a portion of unpredictability. 
In this manner, it is impossible to know how many particles will be produced by a cell
until the infection takes place and the progeny is released. 
Thus, fitness distributions provide a simple way to accommodate this unpredictability 
into each viral replication cycle.
In fact, as shown here, some types of fitness distributions may have a substantial
impact on the evolution of the viral population, most notably the power law. 
The extreme behavior produced by the power law resembles that of the ``viral load blips''
frequently observed in HIV patients under highly active antiretroviral therapy
(HAART)~\cite{DML03,NKK05,NK06,LKSN06,G07,RP09a,RP09b}, with undetectable or very low 
viral loads.  
This particular prediction of the model agrees with the assumption that these events 
are simply due to random fluctuation of the replication process, since after a ``blip'' 
the viral load quickly returns to its basal values. 

The second main result of this paper is concerned with the mechanisms that drive a 
RNA virus population to extinction.
As mentioned before, in the framework of multitype branching processes, there are 
essentially to main attributes associated with this type of event: the probability 
of occurrence of deleterious effects $d$ and the maximum replicative capability $R$.
If in addition to these two, one also considers the carrying capacity as a fundamental
parameter of the phenotypic model then it is possible to show that the two principal 
mechanisms of extinction, lethal mutagenesis and mutational meltdown, are based on 
the same mathematical principle.
Therefore, as far as the phenotypic model is concerned, this is a proof of the
claim~\cite{MOBJ17} that these two mechanisms are ``two sides of the same coin''.

Finally, for the sake of simplicity the phenotypic model considers only three basic 
types of fitness effects; the deleterious, beneficial and neutral.
However, fitness effects represent a broad group forces acting on virus replication 
and one possible direction for further investigation would be to ungroup some of these 
forces and test their action.
For example, on the deleterious side, the inclusion of defective interfering particles
could yield another extinction mechanism, whereas on the beneficial side, the
inclusion of recombination could help the viral population escape from extinction. 

\paragraph{Acknowledgments.}
LG acknowledges the support of FAPESP through the grant number 14/13382-1. 
BG and DC received financial support from CAPES.

\paragraph{Software Availability and Requirements.}
The \texttt{ENVELOPE} program was written in \texttt{C++} programming language, 
using the \texttt{Qt 4.8.6} framework, with the \texttt{Qwt 5.2.1} library. 
It runs on \texttt{Linux} and \texttt{MAC-OSX} operating systems and requires at 
least 2 GB of RAM memory and 1.5 MB of disk space.
Its distribution is free to all users under the LGPL license.
Binary files for \texttt{Linux} and \texttt{MAC-OSX} operating systems are available 
for download at:
\href{https://envelopeviral.000webhostapp.com}%
{\texttt{https://envelopeviral.000webhostapp.com}}

\paragraph{Author contribution:} 
LG and DC contributed equally to this work.
LMRJ and FA contributed equally to this work.
Conceived the model and formulated the underlying theory: LMJR and FA.
Implemented the software: LG, DC and BG. 
Simulated the model and analyzed the output: LG and DC.
Wrote the paper: LMRJ and FA.


\addcontentsline{toc}{section}{References}

\providecommand{\noopsort}[1]{}\providecommand{\singleletter}[1]{#1}%


\appendix

\section*{Appendices}

\section{Review of Multitype Branching Process Theory}
\label{sec:APP1}

A \emph{discrete-time multitype branching process} with \emph{types} or \emph{classes}
indexed by a non-negative integer $r$ ranging from $0$ to $R$ is described by a 
sequence of vector-valued random variables
$\bm{Z}_n=(Z_n^0,\ldots,Z_n^R)$, ($n=0,1,\ldots$), where $Z_n^r$ is the number of
particles of type or class $r$ in the $n$-th generation.
The initial population is represented by a vector of non-negative integers $\bm{Z}_0$ 
(also called a \emph{multi-index}) which is non-zero and non-random.
The time evolution of the population is determined by a vector-valued discrete
probability distribution $\bm{\zeta}(\bm{i})=\big(\zeta_r(\bm{i})\big)$, defined on the 
set of multi-indices $\bm{i}=(i^0,\ldots,i^R)$, called the \emph{offspring distribution} 
of the process, which is usually encoded as the coefficients of a vector-valued 
multivariate power series $\bm{f}(\bm{z})=\big(f_r(\bm{z})\big)$, called 
\emph{probability generating function} (PGF).

The \emph{mean matrix} or the \emph{matrix of first moments} $\bm{M}=\{M_{ij}\}$ of a
multitype branching process describes how the average number of particles in each 
type or class evolves in time and is defined by $M_{ij}=\Expec(Z_1^i|Z^j_0=1)$,
where $Z^j_0=1$ is the abbreviation of $\bm{Z}_0=(0,\ldots,1,\ldots,1)$.
In terms of the probability generating function $\bm{f}=(f_0,\ldots,f_R)$ 
it is given by
\begin{equation} \label{eq:MMFORMULA}
 M_{ij}=\dfrac{\partial f_j}{\partial z_i}(\bm{s})\bigg|_{\bm{s}=\bm{1}}
\end{equation}
where $\bm{1}=(1,1,\ldots,1)$.
Typically, the mean matrix $\bm{M}$ is non-negative and hence it has a 
largest non-negative eigenvalue. 
When the largest eigenvalue is positive, it coincides with the spectral 
radius of $\bm{M}$ and it is called, following Kimmel and Axelrod~\cite{K02}, 
the \emph{malthusian parameter} $\mu$.

The vector of \emph{extinction probabilities} of a multitype branching process, 
denoted by $\bm{\gamma}=(\gamma_0,\ldots,\gamma_R)$, where 
$0 \leqslant\gamma_r\leqslant 1$, is defined by the condition that $\gamma_r$ is 
the probability that the process eventually become extinct given that initially 
there was exactly one particle of class $r$.

The classification theorem of multitype branching proceses states that there are 
only three possible regimes for a multitype branching process~\cite{H63,AN72,K02}:
\begin{description}
\item[\textbf{Super-critical:}] If $\mu>1$ then $0\leqslant\gamma_r<1$ for all $r$ 
      and, with positive probability the population survives indefinitely.
\item[\textbf{Sub-critical:}] If $\mu<1$ then $\gamma_r=1$ for all $r$ and with 
      probability $1$ the population becomes extinct in finite time. 
\item[\textbf{Critical:}] If $\mu=1$ then $\gamma_r=1$ for all $r$ and with 
      probability $1$ the population becomes extinct, however, the expected 
      time to the extinction is infinite.     
\end{description}

When a multitype branching process is super-critical it is expected that, 
according to the ``Malthusian Law of Growth'' it will grow indefinitely at a
geometric rate proportional to $\mu^n$, where $\mu$ is the malthusian parameter,
$\bm{Z}_n \approx \mu^n \,\bm{W}_n$ for some bounded random vector $\bm{W}_n$,
when $n \to \infty$.
The formalization of the above heuristic reasoning is given by the Kesten-Stigum limit 
theorem for super-critical multitype branching processes (see~\cite{KS66a,KS66b,KS67}).
If $\bm{W}_n=\bm{Z}_n/\mu^n$ then there exists a scalar random variable $W \neq 0$ 
such that, with probability one,
\begin{equation} \label{eq:KS1}
 \lim_{n\to\infty} \bm{W}_n = W \,\bm{u}
\end{equation}
where $\bm{u}$ is the right eigenvector corresponding to the malthusian parameter 
$\mu$ and
\begin{equation} \label{eq:KS2}
 \Expec(W|\bm{Z}_0)=\bm{v}^{\mathrm{t}} \bm{Z}_0
\end{equation}
where $\bm{v}$ is the left eigenvector corresponding to the malthusian parameter $\mu$.
The vectors $\bm{u}$ and $\bm{v}$ may be normalized so that
$\bm{v}^{\mathrm{t}}\bm{u}=1$ and $\bm{1}^{\mathrm{t}}\bm{u}=1$
where ${}^{\mathrm{t}}$ denotes the transpose of a vector.
Moreover, under the assumption that $\bm{M}$ is non-negative (which is satisfied 
by the phenotypic model \eqref{eq:MEANGENERAL}), the right and left eigenvectors 
corresponding to the malthusian parameter are non-negative.

The normalization of right eigenvector $\bm{u}=(u_0,\ldots,u_R)$ implies that
$\sum_r u_r=1$ and therefore one has the ``law of convergence of types'' 
(see~\cite{KLPP94})
\begin{equation} \label{eq:CONVTYPES}
 \lim_{n\to\infty} \dfrac{\bm{Z}_n}{|\bm{Z}_n|} = \bm{u} \,,
\end{equation}
where $|\bm{Z}_n|=\sum_r Z_n^r$ is the \emph{total population} at the $n$-th generation
and the equality holds almost surely.
Equation~\eqref{eq:CONVTYPES} asserts that the asymptotic proportion of a replicative
class $r$ converges almost surely to the constant value $u_r$.

In particular, equation~\eqref{eq:CONVTYPES} implies that the malthusian parameter
is the \emph{asymptotic relative growth rate} of the population
\begin{equation} \label{eq:MPAV}
 \mu = \lim_{n\to\infty} \dfrac{|\bm{Z}_{n}|}{|\bm{Z}_{n-1}|} 
     = \lim_{n\to\infty} \dfrac{1}{|\bm{Z}_{n-1}|} \,\sum_{j=1}^{|\bm{Z}_{n-1}|} \#[\,j\,]
\end{equation}
since $|\bm{Z}_{n-1}|$ may be interpreted as the set of ``parental particles'' of the
particles in the $n$-th generation and $|\bm{Z}_{n}|$ is the sum of the ``progeny sizes'' 
$\#[\,j\,]$ of the ``parental particles'' $j$ from the previous generation.

Now consider the quantitative random variable $\varrho$ defined on the set of
classes $\{0,\ldots,R\}$ and having probability distribution $(u_0,\ldots,u_R)$, 
called the \emph{asymptotic distribution of classes}.
When the classes are indexed by their expectation values the variable $\varrho$ 
associates to a random particle its expected class
\[
 \Prob(\varrho=r)=u_r \,.
\]
Therefore, one can define the \emph{average reproduction rate} of the population as
\begin{equation} \label{eq:ARR}
 \langle \varrho \rangle  = \sum_{r=0}^R r \, u_r \,.
\end{equation}
Using equations~\eqref{eq:KS1}, \eqref{eq:KS2}, \eqref{eq:CONVTYPES} one can
show that the average reproduction rate is equal to the malthusian parameter:
\begin{equation} \label{eq:MPARR}
 \langle\varrho\rangle = \mu \,.
\end{equation}
The \emph{average population size} at the $n$-th generation is
$|\langle \bm{Z}_n \rangle| = \sum_{r=0}^R \langle Z^r_n \rangle$.
Then for $n\to\infty$, equation~\eqref{eq:KS1} gives
$|\langle \bm{Z}_n \rangle| \approx \mu^n |\langle \bm{W}_n \rangle| \approx
\mu^n \langle W \rangle$ and so
\begin{equation} \label{eq:MUEST}
 \mu = \lim_{n\to\infty}\dfrac{|\langle \bm{Z}_{n} \rangle|}{|\langle \bm{Z}_{n-1} \rangle|}
\end{equation}
On the other hand, from the definition of mean matrix and its 
form~\eqref{eq:MEANGENERAL}, one has
\[
 |\langle \bm{Z}_n \rangle| = |\bm{M}\,\langle \bm{Z}_{n-1} \rangle|
 =\sum_{r=0}^R r\,\langle Z^r_{n-1} \rangle \,.
\]
Now dividing by $|\langle \bm{Z}_{n-1} \rangle|$ and taking the limit $n\to\infty$ 
gives
\[
 \mu = \lim_{n\to\infty}\dfrac{|\langle \bm{Z}_{n} \rangle|}{|\langle \bm{Z}_{n-1} \rangle|}
 = \lim_{n\to\infty}\sum_{r=0}^R r \,\dfrac{\langle Z^r_{n-1}\rangle}{|\langle
   \bm{Z}_{n-1} \rangle|}
 = \sum_{r=0}^R r \, u_r = \langle \varrho \rangle
\]
where here we used equations~\eqref{eq:KS2} and \eqref{eq:CONVTYPES} in the third
equality from left to right.

In analogy with the characterization of the malthusian parameter as given
by equation~\eqref{eq:MPAV}, one may define the \emph{asymptotic populational variance}
\begin{equation} \label{eq:APV}
 \sigma^2 = \lim_{n\to\infty} \dfrac{1}{|\bm{Z}_{n-1}|} \,
 \sum_{j=1}^{|\bm{Z}_{n-1}|} \#[\,j\,]^2 - \mu^2
\end{equation}
and in analogy with the \emph{mean reproduction rate}, one may define the (squared)
\emph{phenotypic diversity} as
\begin{equation} \label{eq:PD}
 \sigma_{\varrho}^2 = \langle \varrho^2 \rangle - \langle \varrho \rangle^2
\end{equation}

By decomposing the sum in equation~\eqref{eq:APV} according to the classes $r$, 
one obtains
\[
 \sum_{j=1}^{|\bm{Z}_{n-1}|} \#[\,j\,]^2 = \sum_{r=0}^R \sum_{j_r=1}^{Z_{n-1}^r} \#[\,j_r\,]^2
\] 
where $j_r$ runs over the particles of class $r$ for $r=0,\ldots,R$ and
$\#[\,j_r\,]$ are independent random variables assuming non-negative values
with probability distribution $t_r$, called \emph{fitness distribution} of class $r$.

Denoting the variance of the fitness distribution $t_r$ by $\sigma^2_{r}$,
one may write the limit in equation~\eqref{eq:APV} as
\[
\begin{split}
 \sigma^2 
 & = \lim_{n\to\infty} \dfrac{1}{|\bm{Z}_{n-1}|} \,
     \sum_{j=1}^{|\bm{Z}_{n-1}|} \#[\,j\,]^2-\mu^2 \\
 & = \lim_{n\to\infty} \dfrac{1}{|\bm{Z}_{n-1}|} \,\sum_{r=0}^R
     \left[Z^r_{n-1} \left(\dfrac{1}{Z^r_{n-1}} 
     \sum_{j_r=1}^{Z_{n-1}^r}\#[\,j\,]^2 - r^2 \right)+Z^r_{n-1}\right]-\mu^2 \\
 & = \lim_{n\to\infty} \dfrac{1}{|\bm{Z}_{n-1}|} \,\sum_{r=0}^R
     (\sigma^2_{r}+r^2)Z^r_{n-1}-\mu^2  
\end{split}
\]
Then equations~\eqref{eq:CONVTYPES}, \eqref{eq:MPARR} and~\eqref{eq:PD} give
\begin{equation} \label{eq:APVDELTA}
 \sigma^2 = \sum_{r=0}^R (\sigma^2_{r}+r^2 )\,u_r-\mu^2 
          = \sum_{r=0}^R \sigma^2_{r}\,u_r+\sigma^2_{\varrho}
\end{equation}
The difference between the asymptotic populational variance and the (squared)
phenotypic diversity, called \emph{normalized populational variance}, is the 
weighted average of the variances of the fitness distributions
\begin{equation}
 \phi = \sigma^2 - \sigma^2_{\varrho} = \sum_{r=0}^R \sigma^2_{r}\,u_r \,.
\end{equation}
In particular, when the family of fitness distributions is the deterministic family 
the populational variance is exactly the phenotypic diversity (that is $\phi=0$).
This is an expected result since the Delta distributions $t_r(k)=\delta_{rk}$ have 
zero variance and hence the only source of fluctuation of the population size is due 
to its stratification into replicative classes, which is expressed by the phenotypic
diversity.

\section{Mathematical Basis of the Phenotypic Model}
\label{sec:APP0}

Based on the general aspects of the phenomenon of viral replication described before 
it is compelling to model it in terms of a branching process. 
At each replicative cycle, every parental particle in the replicative class $r$ produces
a random number of progeny particles that is independently drawn from the corresponding
fitness distribution.

A \emph{fitness distribution} is a member of a location-scale family of discrete 
probability distributions $t_r$ parameterized by the replicative classes  
($r=0,\ldots,R$) assuming non-negative integer values and normalized so that the
expectation value of $t_r$, defined as $\sum_{k} k\,t_r(k)$, is exactly $r$ and 
$t_0(k)=\delta_{k0}$. 
Here $\delta_{kr}=1$ if $k=r$ and $\delta_{kr}=0$ if $k \neq r$.
Therefore, each particle in the viral population is characterized by the mean value 
of its fitness distribution, called \emph{mean replicative capability}.
Viral particles with replicative capability equal to zero ($0$) do not generate
progeny; viral particles with replicative capability one ($1$) generate one particle
on average; viral particles with replicative capability two ($2$) generate two
particles on average, and so on.
Typical examples of location-scale families of discrete probability distributions 
that can be used as fitness distributions are:
\begin{enumerate}[(a)]
\item The family of \emph{Deterministic (Delta) distributions}: $t_r(k)=\delta_{kr}$.
\item The family of \emph{Poisson distributions}: $t_r(k)=\ee^{-r}\tfrac{r^k}{k!}$.
\end{enumerate}
Note that in the first example, the replicative capability is completely concentrated 
on the mean value $r$ -- that is, the particles have deterministic fitness.
On the other hand, in the second example the fitness is truly stochastic.

\pagebreak

During the replication, each progeny particle always undergoes one of the following
effects: 
\begin{description}
\item[\textbf{Deleterious effect:}] the mean replication capability of the respective 
      progeny particle decreases by one. 
      Note that when the particle has capability of replication equal to $0$ it will 
      not produce any progeny at all.
\item[\textbf{Beneficial effect:}] the replication capability of the respective progeny 
      particle increases by one. 
      If the mean replication capability of the parental particle is already the maximum 
      allowed then the mean replication capability of the respective progeny particles 
      will be the same as the replicative capability of the parental particle. 
\item[\textbf{Neutral effect:}] the mean replication capability of the respective 
      progeny particle remains the same as the mean replication capability of the 
      parental particle.
\end{description}
To define which effect will occur during a replication event, probabilities 
$d$, $b$ and $c$ are associated, respectively, to the occurrence of \emph{deleterious}, 
\emph{beneficial} and \emph{neutral} effects.
The only constraints these numbers should satisfy are $0\leqslant d,b,c\leqslant 1$
and $b+c+d=1$.
In the case of \emph{in vitro} experiments with homogeneous cell populations the
probabilities $c$, $d$ and $b$ essentially refer to the occurrence of mutations.

The probability generating function (PGF) of the phenotypic model with $b=0$ and 
$t_r(k)=\delta_{kr}$ is (see Antoneli \etal~\cite{ABCJ13b,ABCJ13a} for details):
\begin{equation} \label{eq:GENFUNC}
\begin{split}
 f_0(z_0,z_1,\ldots,z_R) & = 1 \\
 f_1(z_0,z_1,\ldots,z_R) & = dz_0+cz_1 \\
 f_2(z_0,z_1,\ldots,z_R) & = (dz_1+cz_2)^2 \\
                        & \vdots \\
 f_R(z_0,z_1,\ldots,z_R) & = (dz_{R-1}+cz_R)^R
\end{split}
\end{equation}
Note that the functions $f_r(z_0,z_1,\ldots,z_R)$ are polynomials whose coefficients 
are exactly the probabilities of the binomial distribution $\mathrm{binom}(k;r,1-d)$.
The PGF in the case with general beneficial effects and with a general family of 
fitness distribution (which reduces to the previous PGF when $b=0$ and 
$t_r(k)=\delta_{kr}$) is given by.
\begin{equation} \label{eq:GENFUNCG}
\begin{split}
 f_0(z_0,z_1,\ldots,z_R) & = 1 \\
 f_1(z_0,z_1,\ldots,z_R) & = \sum_{k=0}^\infty \,t_1(k)\, (dz_0+cz_1+bz_2)^k \\
 f_2(z_0,z_1,\ldots,z_R) & = \sum_{k=0}^\infty \,t_2(k)\, (dz_1+cz_2+bz_3)^k \\
                        & \vdots \\
 f_R(z_0,z_1,\ldots,z_R) & = \sum_{k=0}^\infty \,t_R(k)\, (dz_{R-1}+(c+b)z_R)^k
\end{split}
\end{equation}
Note that in the last equation the beneficial effect acts like the neutral effect.
This is a kind of ``consistency condition'' ensuring that the populational replicative 
capability is, on average, upper bounded by $R$. 
Even though it is possible that a parental particle in the replicative classes $R$ 
eventually has more than $R$ progeny particles when $t_r$ is not deterministic,
the average progeny size is always $R$.

Finally, it is easy to see that the PGF of the two-dimensional case of the phenotypic 
model with $b=0$ and $z_0=1$ (and ignoring $f_0$) reduces to
\begin{equation} \label{eq:DSS}
  f(z)~=~\sum_{k=0}^\infty \,t(k)\, ((1-c)+cz)^k~=~\sum_{k=0}^\infty \,t(k)\, (1-c(1-z))^k \,.
\end{equation}
This is formally identical to the PFG of the single-type model proposed 
by~\cite[p. 255, eq. (49)]{DSS85} for the evolution of polynucleotides. 
In their formulation, $c=p^\nu$ is the probability that a given copy of a 
polynucleotide is exact, where the polymer has chain length of $\nu$ nucleotides 
and $p$ is the probability of copying a single nucleotide correctly. 
The \emph{replication distribution} $t(k)$ provides the number of copies a
polynucleotide yields before it is degraded by hydrolysis.

A remarkable property of the phenotypic model that was fully explored in 
Antoneli \etal~\cite{ABCJ13b,ABCJ13a} is the fact that when $b=0$ the phenotypic
model is ``exactly solvable'' in a very specific sense.

It is straightforward form the generating function \eqref{eq:GENFUNCG}, using
formula~\eqref{eq:MMFORMULA}, that the matrix of the phenotypic model is given by
\begin{equation} \label{eq:MEANGENERAL}
\bm{M}=\begin{pmatrix}
 0 & d &  0 &  0 &  0 & \ldots & 0 \\
 0 & c & 2d &  0 &  0 & \ldots & 0 \\
 0 & b & 2c & 3d &  0 & \ldots & 0 \\
 0 & 0 & 2b & 3c & 4d & \ldots & 0 \\
 0 & 0 &  0 & 3b & 4c & \ldots & 0 \\
\vdots & \vdots & \vdots & \vdots & \vdots & \ddots & Rd \\
 0 & 0 &  0 & 0 & 0 & (R-1)b & R(c+b)
\end{pmatrix} \,.
\end{equation}
Note that the mean matrix does depend on the fitness distributions $t_r$ only through
their mean values, since $t_r$ are normalized to have the mean value $r$.

Assume for a moment that $b=0$ (hence $c=1-d$).
Then the mean matrix becomes upper-triangular and hence its eigenvalues are the 
diagonal entries $\lambda_r=r(1-d)$ and the malthusian parameter $\mu$ is the 
largest eigenvalue $\lambda_R$:
\begin{equation} \label{eq:MALTHUSIAN}
 \mu=R(1-d) \,.
\end{equation}
Now suppose that $b \neq 0$ is small compared to $d$ and $c$ (hence $c=1-d-b$).
Then spectral perturbation theory allows one to write the malthusian parameter 
$\mu$ as a power series
\[
 \mu = \mu_0 + \mu_1 b + \mu_2 b^2 + \cdots
\]
where $\mu_0$ is the malthusian parameter for the case $b=0$ and $\mu_j$ are
functions of the form $R\,\tilde{m}_j(d,R)$.
A lengthy calculation (see~\cite{ABCJ13a}) gives the following result:
\begin{equation} \label{eq:PERTEXPAN}
 \mu = R \left((1-d) + (R-1)\dfrac{d}{1-d}\,b + \bigO(b^2)\right) \,.
\end{equation}

Let us return to the case $b=0$ and consider the eigenvectors corresponding 
to the malthusian parameter $\mu$.
The right eigenvector $\bm{u}=(u_0,\ldots,u_R)$ and the left eigenvector 
$\bm{v}=(v_0,\ldots,v_R)$ may be normalized so that $\bm{v}^{\mathrm{t}}\bm{u}=1$ and
$\bm{1}^{\mathrm{t}}\bm{u}=1$, where ${}^{\mathrm{t}}$ denotes the transpose of a vector.
In~\cite{ABCJ13a} it is shown that the normalized right eigenvector 
$\bm{u}=(u_0,\ldots,u_R)$ is given by
\begin{equation} \label{eq:REIGENVEC}
 u_r=\binom{R}{r} \, (1-d)^r \, d^{R-r} \,.
\end{equation}
The fact that $\bm{u}$ is a binomial distribution is not accidental.
Indeed, it can be shown that $\bm{u}$ is the probability distribution of
a quantitative random variable $\varrho$ defined on the set of replicative classes 
$\{0,\ldots,R\}$, called the \emph{asymptotic distribution of classes}, such that
$u_r=\mathrm{binom}(r;R,1-d)$ gives the limiting proportion of particles in the 
$r$-th replicative class.
Finally, when $b \neq 0$ is small, spectral perturbation theory ensures that
\begin{equation} \label{eq:REIGENVECPERT}
 u_r=\binom{R}{r} \, (1-d)^r \, d^{R-r} + \bigO(b) \,.
\end{equation}

\begin{figure}[ht] 
\begin{center}
 \includegraphics[scale=0.5]{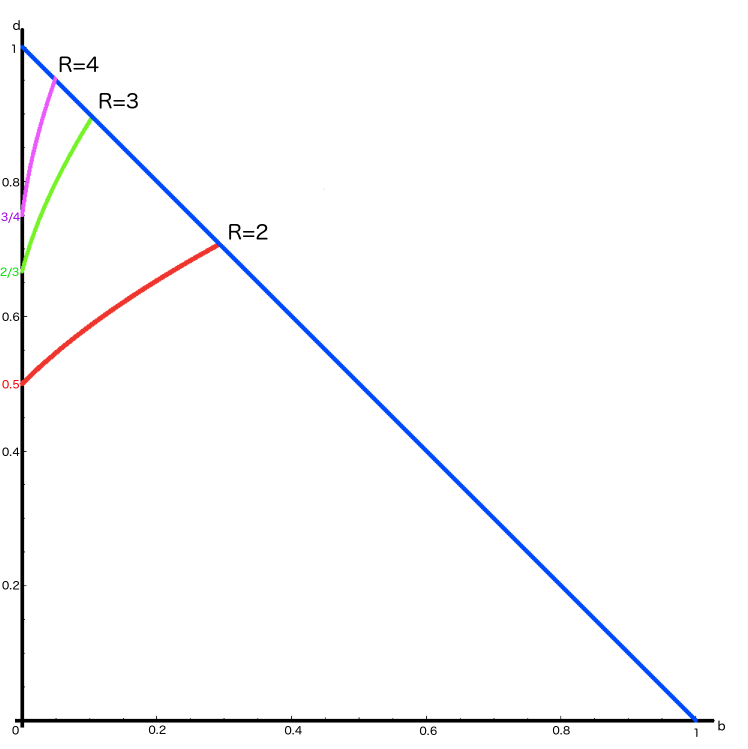}
 \caption{\label{fig:simplex} Parameter space of the phenotypic model.
 The blue line is boundary $b+d=1$.
 The red, green and magenta curves are the critical curves $\mu(b,d,R)=1$
 for $R=2,3,4$, respectively.} 
\end{center}
\end{figure}

The phenotypic model is completely specified by the choice of the two probabilities 
$b$ and $d$ (since $c=1-b-d$), the maximum replicative capability $R$ and a choice 
of a location-scale family of fitness distributions.
Independently of the choice of family of fitness distributions the parameter space 
of the model is the set $\triangle^2 \times \{R\in\N:R \geqslant 1\}$,
where $\triangle^2=\{(b,d)\in [0,1]^2: b+d \leqslant 1\}$ is the 
\emph{two-dimensional simplex} (see Figure~\ref{fig:simplex}).

In this parameter space one can consider the \emph{critical curves} 
$\mu(b,d,R)=1$, where $\mu(b,d,R)$ is the malthusian parameter as a function 
of the parameters of the phenotypic model.
For each fixed $R$, the corresponding \emph{critical curve} is independent of 
the fitness distributions and represents the parameter values $(b,d)$ such that 
the branching process is critical.
Moreover, each curve splits the simplex into two regions representing the parameter
values where the branching process is super-critical (above the curve) and
sub-critical (below the curve).

One of the main results of~\cite{ABCJ13a} is a proof of the lethal mutagenesis
criterion~\cite{BSW07} for the phenotypic model, provided one assumes that all
fitness effects are of a purely mutational nature.
Recall that~\cite{BSW07} assumes that all mutations are either neutral or deleterious
and consider the \emph{mutation rate} $U=U_d+U_c$, where the component $U_c$ comprises 
the purely neutral mutations and the component $U_d$ comprises the mutations with a 
deleterious fitness effect.
Furthermore, $R_{\mathrm{max}}$ denotes the maximum replicative capability among all
particles in the viral population. 
The \emph{lethal mutagenesis criterion} proposed by~\cite{BSW07} states that a 
sufficient condition for extinction is 
\begin{equation} \label{eq:LETHALMUT}
 R_{\mathrm{max}}\,\ee^{-U_d} < 1 \,.
\end{equation}
According to~\cite{BSW07,BSW08}, $\ee^{-U_d}$ is both the mean fitness level and also 
the fraction of offspring with no non-neutral mutations.
Moreover, in the absence of beneficial mutations and epistasis~\cite{KM66} the only 
type of non-neutral mutations are the deleterious mutations.
Therefore, in terms of fitness effects, the probability $\ee^{-U_d}$ corresponds to 
$1-d=c$.
Since the evolution of the mean matrix depends only on the expected values of the
fitness distribution $t_r$, it follows that $R_{\mathrm{max}}$ corresponds to $R$.
That is, the lethal mutagenesis criterion of \eqref{eq:LETHALMUT} is formally 
equivalent to \emph{extinction criterion}
\begin{equation} \label{eq:LETHALMUTALT}
 R(1-d) < 1
\end{equation}
which is exactly the condition for the phenotypic model to become \emph{sub-critical}.
Formula \eqref{eq:PERTEXPAN} for the malthusian parameter provides a generalization 
of the extinction criterion \eqref{eq:LETHALMUTALT} without the assumption that
that all effects are either neutral or deleterious. 
If $b>0$ is sufficiently small (up to order $\bigO(b^2)$) and
\begin{equation} \label{eq:LETHALMUTINTERACT}
 R \left((1-d) + (R-1)\dfrac{bd}{1-d}\right) < 1
\end{equation}
then, with probability one, the population becomes extinct in finite time.

On the other hand, a deeper exploration of the implications of non-zero 
beneficial effects allowed for the discovery of a \emph{non-extinction criterion}.
If $b>0$ is sufficiently small (up to order $\bigO(b^2)$), $R$ is sufficiently large 
($R \geqslant 10$ is enough) and
\begin{equation} \label{thm:NOEXTINCTAPPROX}
 R^3 \, b > 1
\end{equation}
then, asymptotically almost surely, the population can not become extinct 
by increasing the deleterious probability $d$ towards its maximum value $1-b$ 
(see~\cite{ABCJ13a} for details).
In other words, a small increase of the beneficial probability may have a drastic
effect on the extinction probabilities, possibly rendering the population impervious
to become extinct by lethal mutagenesis (i.e., by increase of deleterious effects).

In the theory of multitype branching processes there several variations as follows:
continuous time, age dependent, self-regulated, etc, (see~\cite{AN72,H63,K02}).
The implementation of a variation of the theory of multitype branching process
accounting for the notions of evolutionary entropy and directionality theory
(see~\cite{D05,D2013}) could be useful for studies on viral evoltution. 
In this case, the malthusian parameter $\mu$, which is the dominant
eigenvalue of the mean matrix, could be expressed as the sum of two terms
\[
 \mu = H + \Phi \,.
\]
The quantity $H$ is called \emph{evolutionary entropy} and $\Phi$ is called
the \emph{reproductive potential}~\cite{D2013}. 
An interesting direction to follow would be to develop an extinction criterion
based on evolutionary entropy instead of the malthusian parameter.

\section{The Deterministic Selection Equation}
\label{sec:APP1/2}

According to~\cite{DSS85,D1985,D1987}, one may associate to a multitype branching
process a system of difference (or ordinary differential) equations, called 
\emph{selection equations}, on the space of discrete probability distributions 
$\triangle^{R+1}=\{\bm{p}\in\R^{R+1}:p_j\geqslant 0;\sum_{j}p_j=1\}$ over the 
finite state set $\{0,\ldots,R\}$.
Given a discrete multitype branching process $\bm{Z}_n$, then the expectation 
values $\langle\bm{Z}_n\rangle$ satisfy $\langle\bm{Z}_n\rangle=\bm{M}^n\bm{Z}_0$, 
with $\bm{M}$ being the mean matrix of $\bm{Z}_n$. 
Hence $\bm{Z}_n$ is given by iteration of the difference equation 
$\bm{z}_n= \bm{M}\bm{z}_{n-1}$.
This yields a discrete-time selection equation by normalizing the difference
equation, thereby obtaining
\begin{equation} \label{eq:DSE}
 \bm{x}_n = \dfrac{1}{\bm{1}^\transp\bm{M}\bm{x}_{n-1}}\bm{M}\bm{x}_{n-1}
\end{equation}
where $\bm{1}=(1,\ldots,1)$. 
Then, passing~\eqref{eq:DSE} to continuous time one obtains a continuous-time 
selection equation
\begin{equation} \label{eq:GSE}
 \dot{\bm{x}} = [\bm{M}\bm{x}-\bm{x}(\bm{1}^\transp\bm{M}\bm{x})]
 \dfrac{1}{\bm{1}^\transp\bm{M}\bm{x}} \,.
\end{equation}
Multiplying the right hand side of equation~\eqref{eq:GSE} with the factor 
$\bm{1}^\transp\bm{M}\bm{x}$, which is always strictly positive on $\triangle^{R+1}$, 
corresponds to a change in velocity (re-scaling time) and so, the solutions 
of~\eqref{eq:GSE} are the same as the solutions of 
\begin{equation} \label{eq:ESE}
 \dot{\bm{x}} = \bm{M}\bm{x}-\bm{x}(\bm{1}^\transp\bm{M}\bm{x})
\end{equation}
It follows from general considereations (see~\cite{DSS85,D1985,D1987}) that
equation~\eqref{eq:ESE} has a unique global stable equilibrium on $\triangle^{R+1}$ 
given by the normalized right eigenvector $\bm{u}$ of $\bm{M}$ corresponding to its 
largest eigenvalue $\mu$.
In this sense, the deterministic selection equation yields a description of the
evolution of the normalized mean values of the corresponding stochastic model,
thus definig a \emph{mean field (macroscopis) dynamics} representing the infinite 
population limit of the branching process.

\section{The Power Law Distribution Family}
\label{sec:APP2}

It is typical to parameterize power law distributions by the \emph{exponent} 
$s$, which measures the ``weight of the tail'' of the distribution.
However, we need to have a location-scale parameterized family in order to 
impose the same normalization as we have done for the other types of distributions.
Therefore, we define the \emph{power law} distribution with \emph{mean value} 
$r$ by
\[
 \mathfrak{z}_r(k) = \frac{(k-1)^{s(r)}}{\zeta(s(r))}
\]
for $k=0,1,\ldots,\infty$ and $r \geqslant 1$, where $\zeta(s)$ 
is the \emph{Riemann zeta function}, defined for $s>1$, by
\[
 \zeta(s) = \sum_{n=1}^{\infty} \frac{1}{n^s}
\]
and the function $s(r)$ is given by the inverse function of
\[
 r = \varphi(s) = \frac{\zeta(s-1)}{\zeta(s)}-1 \,.
\]
Namely, $s=\varphi^{-1}(r)$ for $r\geqslant 1$ and hence when
$1 \leqslant r < \infty$ the exponent $s$ satisfies $3<s<2$.
Moreover, the \emph{Laurent} series expansion for $r\to\infty$ 
($s \to 2$) is given by:
\begin{equation} \label{eq:MEANFROMEXP}
 s(r) \approx 2 + \frac{6}{\pi^2(1+r-C)} \,. 
\end{equation}
The constant $C$ in the previous formula is given by
$C = [6 \gamma\pi^2 - 36\,\zeta'(2)]/\pi^4 \approx 0.6974$,
where $\gamma$ is \emph{Euler's constant} and $\zeta'(2)$ is
the derivative of $\zeta(s)$ evaluated at $2$.
Observe that when the mean value $r \geqslant 1$, the exponent 
$s<3$, and so the variance of $\mathfrak{z}_r(k)$ is infinite.

The implementation of the pseudo-random generation of samples from
the distribution $\mathfrak{z}_r(k)$ in the \texttt{ENVELOPE} program
is based on the algorithm of~\cite{D1986} for the \emph{Zipf distribution}
on the positive integers, using formula~\ref{eq:MEANFROMEXP} for the
computation of the exponent $s$ given the mean value $r$.
Pseudo-random generation for the remaining fitness distributions were 
implemented using the standard library of \texttt{C++} programing language 
(this library requires \texttt{C++} (2011) or superior).

\section{Main Routines of the ENVELOPE Program}
\label{sec:APP3}

\begin{algorithm}
\begin{algorithmic}[1]
\State $\triangleright$ \texttt{Variables Defined by the User}
\State Real $b,d$; \Comment {\texttt{Beneficial and Deleterious Probabilities}}
\State Integer $R$; \Comment {\texttt{Maximum Replicative Cabability}}
\State Integer $N$; \Comment {\texttt{Maximum Generation Time}}
\State Integer $K$; \Comment {\texttt{Maximum Numer of Particles}}
\State Intehre $type$; \Comment{\texttt{Type of Fitness Distribution}}
\State Integer Vector $initial\_population[0,\ldots,R]$;
               \Comment {\texttt{Initial Particle Distribution}}
\end{algorithmic}
\end{algorithm}

\begin{algorithm}
\begin{algorithmic}[1]
\State $\triangleright$ \texttt{Global Variables}
\State Integer Vector $malthusian[0,\ldots,N]$;
               \Comment {\texttt{Malthusian Parameter per Generation}}
\State Real Matrix $class\_distribution[0,\ldots,R][0,\ldots,N]$;
               \Comment {\texttt{Class Distribution per Generation}}
\State Real Vector $mean\_rho[0,\ldots,N]$;
               \Comment {\texttt{Average Reproduction Rate per Generation}}
\State Real Vector $diversity[0,\ldots,N]$;
               \Comment {\texttt{Phenotypic Diversity per Generation}}
\State Real Vector $entropy[0,\ldots,N]$;
               \Comment {\texttt{Phenotypic Entropy per Generation}}
\end{algorithmic}
\end{algorithm}

\begin{algorithm}
\begin{algorithmic}[1]
\State $\triangleright$ \texttt{Internal Variables}
\State Integer $n$; \Comment {\texttt{Current Generation Time}}
\State Integer $T$; \Comment {\texttt{Current Total Progeney}}
\State Integer $progeny$; \Comment {\texttt{Progeney of a Replicative Class}}
\State Integer $sampled$; \Comment {\texttt{Random Particle Sampled}}
\State Integer Vector $particles[0,\ldots,R]$; 
               \Comment {\texttt{Current Particle Distribution}}
\State Integer Vector $parents[0,\ldots,R]$; 
               \Comment {\texttt{Current Parental Distribution}}
\State Integer Vector $next[0,\ldots,R]$; 
               \Comment {\texttt{Next Generation Particle Distribution}}
\State Real $\textit{effect}$; \Comment {\texttt{Random Number Between $0$ and $1$}}
\end{algorithmic}
\end{algorithm}

\begin{algorithm}
\begin{algorithmic}[1]
\State $\triangleright$ 
       \texttt{Sample from a Fitness Distribution of type $t$ with mean value $m$}
\Function{FitnessDistribution}{Real $m$, Integer $t$}
\State Integer $value$;
\State \textbf{case} $t$ \textbf{do}
\State $t=0$: $value \gets (\textrm{Integer}) \, m$;
\State $t=1$: $value \gets \Call{Poisson}{m}$;
\State $t=2$: $value \gets \Call{Geometric}{1/(1+m)}$;
\State $t=3$: $value \gets \Call{Binomial}{2*m,1/2}$;
\State $t=4$: $value \gets \Call{PowerLaw}{m}$;
\State \textbf{end case}
\State return $value$;
\EndFunction
\end{algorithmic}
\end{algorithm}

\begin{algorithm}
\begin{algorithmic}[1]
\State $\triangleright$ \texttt{Compute the Statistics for the Next Generation}
\Procedure{Statistics}{Integer Vector $u$, Integer Vector $v$, Integer $n$}
\State $malthusian[n] \gets \Call{Sum}{u}/\Call{Sum}{v}$;
\For{$i$ \textbf{from} $0$ \textbf{to} $R$}
\State $class\_distribution[i][n] \gets v[i]/\Call{Sum}{v}$;
\EndFor
\State $mean\_rho[n] \gets \Call{Average}{class\_distribution[n]}$;
\State $diversity[n] \gets \Call{Diversity}{class\_distribution[n]}$;
\State $entropy[n] \gets \Call{Entropy}{class\_distribution[n]}$;
\EndProcedure
\end{algorithmic}
\end{algorithm}

\begin{algorithm}
\begin{algorithmic}[1]
\State $\triangleright$ 
\texttt{Average of Class Distribution}
\Function{Average}{Real Vector $prob$}
\State Real $x=0.0$;
\For{$i$ \textbf{from} $1$ \textbf{to} $R$}
\State $x \gets x + i * prob[i]$;
\EndFor
\State return $x$;
\EndFunction
\end{algorithmic}
\end{algorithm}

\begin{algorithm}
\begin{algorithmic}[1]
\State $\triangleright$ 
\texttt{Diversity of Class Distribution}
\Function{Diversity}{Real Vector $prob$}
\State Real $x=0.0$, $y=0.0$;
\For{$i$ \textbf{from} $1$ \textbf{to} $R$}
\State $x \gets x + i * i * prob[i]$;
\EndFor
\State $y \gets \Call{Average}{prob}$;
\State $x \gets x - y*y$;
\State return $x$;
\EndFunction
\end{algorithmic}
\end{algorithm}

\begin{algorithm}
\begin{algorithmic}[1]
\State $\triangleright$ 
\texttt{Entropy of Class Distribution}
\Function{Entropy}{Real Vector $prob$}
\State Real $x=0.0$;
\For{$i$ \textbf{from} $1$ \textbf{to} $R$}
\State $x \gets x + prob[i] * \Call{Log}{prob[i]}$;
\EndFor
\State return $x$;
\EndFunction
\end{algorithmic}
\end{algorithm}

\begin{algorithm}
\begin{algorithmic}[1]
\State $\triangleright$ \texttt{Initialization}
\State $particles \gets initial\_population$;
\State $n \gets 0$;
\State $\triangleright$ \texttt{Main loop}
\Repeat
\State $T \gets \Call{Sum}{particles}$;
\State $parents \gets particles$;
\State $next \gets [0,\ldots,0]$;
\For{$i$ \textbf{from} $0$ \textbf{to} $R$} \Comment{\texttt{Generate Progeny}}
\State $progeny \gets 0$;
\For{$j$ \textbf{from} $1$ \textbf{to} $parents[i]$}
\State $progeny \gets progeny + \Call{FitnessDistribution}{(\textrm{Real})\,i,type}$;
\EndFor
\State $particles[i] \gets progeny$;
\EndFor
\If {$T <= K$} \Comment {\texttt{Find Replicative Class of Progeny Without Cut Off}}
\For{$i$ \textbf{from} $0$ \textbf{to} $R$}
\For{$j$ \textbf{from} $1$ \textbf{to} $particles[i]$}
\State $\textit{effect} \gets \Call{RandomRealNumber}{0,1}$;
\If {$\textit{effect} <= d$}  \Comment{\texttt{Deleterious Effect}}
\State $next[i-1] \gets next[i-1] + 1$;  
\ElsIf {$\textit{effect} > d+b$}  \Comment{\texttt{Beneficial Effect}}
\State $next[i+1] \gets next[i+1] + 1$;
\Else   \Comment{\texttt{Neutral Effect}}
\State $next[i] \gets next[i] + 1$;
\EndIf
\EndFor
\EndFor
\Else \Comment {\texttt{Find Replicative Class of Progeny With Cut Off}}
\For{$j$ \textbf{from} $1$ \textbf{to} $K$}
\State $sampled \gets \Call{RandomIntegerNumber}{0,T}$;
\For{$i$ \textbf{from} $0$ \textbf{to} $R$}
\If {$sampled < particles[i]$}
\State $T \gets T - 1$;
\State $particles[i] \gets particles[i] - 1$;
\State $\textit{effect} \gets \Call{RandomRealNumber}{0,1}$;
\If {$\textit{effect} <= d$}  \Comment{\texttt{Deleterious Effect}}
\State $next[i-1] \gets next[i-1] + 1$;  
\ElsIf {$\textit{effect} > d+b$}  \Comment{\texttt{Beneficial Effect}}
\State $next[i+1] \gets next[i+1] + 1$;
\Else  \Comment{\texttt{Neutral Effect}}
\State $next[i] \gets next[i] + 1$;
\EndIf
\State break;
\Else
\State $sampled \gets sampled - particles[i]$;
\EndIf
\EndFor
\EndFor
\EndIf
\State $particles \gets next$; \Comment{\texttt{Conclude and Sumarize}}
\State $n \gets n + 1$;
\State \Call{Statistics}{$parents$,$next$,$n$};
\Until {$T = 0$ or $n > N$ or \Call{UserStop}{}};
\end{algorithmic}
\end{algorithm}

\end{document}